\documentclass[fleqn,usenatbib]{mnras}
\usepackage[dvipdfmx]{graphicx}
\usepackage{pdfpages}
\usepackage{amsmath}	% Advanced maths commands
\usepackage{amssymb}	% Extra maths symbols
\usepackage{bm}
\usepackage{multirow}

\title[Impact on non-ideal MHD effects]{Impact of turbulence intensity and fragmentation velocity on dust particle size evolution and non-ideal magnetohydrodynamics effects} 
\author[Y.  Kawasaki et al.]{
  Yoshihiro Kawasaki $^{1}$\thanks{E-mail: kawasaki.yoshihiro.592@s.kyushu-u.ac.jp (YK)}
  and Masahiro N. Machida$^{1}$
  \\
  $^{1}$Department of Earth and Planetary Sciences, Faculty of Sciences, Kyushu University, Fukuoka 819-0395, Japan
}

\date{Accepted XXX. Received YYY; in original form ZZZ}

\pubyear{2015}

\begin{document}
\label{firstpage}
\pagerange{\pageref{firstpage}--\pageref{lastpage}}
\maketitle

\begin{abstract}
We investigate the influence of dust particle size evolution on non-ideal magnetohydrodynamic effects during the collapsing phase of star-forming cores,
taking both the turbulence intensity in the collapsing cloud core and  the fragmentation velocity of dust particles as parameters. 
When the turbulence intensity is small, the dust particles do not grow significantly, and the non-ideal MHD effects work efficiently in high-density regions.
The dust particles rapidly grow in a strongly turbulent environment, while the efficiency of non-ideal MHD effects in such an environment depends on the fragmentation velocity of the dust particles. 
When the fragmentation velocity is small, turbulence promotes coagulation growth and collisional fragmentation of dust particles, producing small dust particles.
In this case, the adsorption of charged particles on the dust particle surfaces becomes efficient and the abundance of charged particles decreases, making non-ideal MHD effects effective at high densities.
On the other hand, when the fragmentation velocity is high, dust particles  are less likely to fragment, even if the turbulence is strong.
In this case, the production of small dust particles  become inefficient and non-ideal MHD effects become less effective.
We also investigate the effect of the dust composition on the star and disk formation processes. 
We constrain the turbulence intensity of a collapsing core and the fragmentation velocity of dust for circumstellar disk formation due to the dissipation of the magnetic field. 
\end{abstract}


%%%%%%%%%%%%%%%%%%%%%%%%%%%%%%%%%%%%%%%%%%%%%%%%%%

%%%%%%%%%%%%%%%%% BODY OF PAPER %%%%%%%%%%%%%%%%%%
\section{Introduction}

Dust particles are an essential ingredient in star and planet formation processes.
Observing the radiation emitted from dust particles provides clues to star-forming activity.
The surfaces of dust particles are production sites for large molecules in molecular clouds \citep{1982A&A...114..245T,1992ApJS...82..167H}. 
Dust particles are also the most fundamental building block of planets.
Regarding non-ideal magnetohydrodynamic (MHD) effects (Ohmic dissipation, ambipolar diffusion, and the Hall effect \citealt{1999MNRAS.303..239W,2002ApJ...573..199N}), 
dust particles absorb charged particles and reduce the degree of ionization in the collapsing cloud core, which enhances the magnetic diffusion coefficients. 
Thus, although the dust-to-gas mass ratio in star-forming regions is as small as about 0.01 \citep{1977ApJ...217..425M}, dust properties such as the particle size and abundance significantly affect the magnetic diffusion coefficients and the dissipation of the magnetic field. 

The magnetic field plays various roles in the star and disk formation processes. 
The angular momentum in the collapsing cloud core is removed by magnetic effects such as magnetic braking \citep[][]{1994ApJ...432..720B,1995ApJ...452..386B,2000ApJ...528L..41T,2005MNRAS.362..369M}, promoting the protostellar mass growth.  
On the other hand, when the magnetic field is well coupled with the gas  (as in ideal-MHD) in a high-density region, the angular momentum transport due to magnetic braking is too efficient to form the circumstellar disk \citep{2003ApJ...599..363A,2008ApJ...681.1356M}. 
Thus, in such a case, the magnetic effect suppresses the disk formation. 
However, it is possible to form the circumstellar disks when considering the dissipation of magnetic field \citep[e.g.][]{2014MNRAS.438.2278M}.
Recently, many circumstellar disks were observed around very young protostars \citep{Tobin2016, 2017ApJ...834..178Y,2020ApJ...902..141S}. 
Thus, it is considered that the dissipation of magnetic field in a high-density or disk-forming region is the most important to understanding the disk formation  \citep{2022arXiv220913765T}.
Therefore, we need carefully consider the dissipation of the magnetic field (or magnetic diffusivities) in the star formation process.

In collapsing cloud cores, the disk is considered to be formed in a magnetically inactive region where the magnetic field is dissipated due to ambipolar diffusion and Ohmic dissipation
\citep{2010A&A...521L..56D, 2010ApJ...724.1006M, 2015ApJ...801..117T, 2016A&A...587A..32M, 2018A&A...615A...5V, 2020ApJ...900..180M, 2021MNRAS.502.4911X, 2021MNRAS.508.2142X}. 
The abundance of charged particles determines the magnetic diffusion coefficients and depends on the dust particle size distribution \citep{2018MNRAS.478.2723Z, 2019MNRAS.484.2119K, 2022ApJ...934...88T}. 
Different particle size distributions give different dust number densities 
and total cross sections related to the adsorption efficiency of charged particles on the surfaces of dust particles.
Therefore, the shape of dust particle size distribution is important when considering non-ideal MHD effects in the formation of a circumstellar disk.

It is crucial to determine how the dust particle size distribution evolves in a collapsing core in order to understand non-ideal MHD effects.
Many past studies used a fixed dust particle size distribution, such as the Mathis, Rumpl, and Nordsieck (MRN) distribution, \citep{1977ApJ...217..425M},
where it is assumed that the dust particle size distribution does not evolve in time when calculating the magnetic diffusion coefficients \citep{2009ApJ...693.1895K,2016A&A...592A..18M, 2016MNRAS.460.2050Z,2018MNRAS.478.2723Z,2019MNRAS.484.2119K, 2020A&A...641A.112L, 2021MNRAS.501.5873W}.
However, dust particles can change their size by collisional coagulation or fragmentation in a dense core 
\citep{2009A&A...502..845O, 2009MNRAS.399.1795H, 2021MNRAS.507.2318V, 2022MNRAS.514.2145B, 2022MNRAS.515.4780T, 2022A&A...658A.191V, 2023A&A...670A..61M}.
A few studies have investigated dust particle size evolution and how it is associated with non-ideal MHD effects.
\citet{2020A&A...643A..17G} showed that dust coagulation growth significantly changes the magnetic diffusion coefficients in a collapsing core.
\citet[][hereafter Paper I]{2022MNRAS.515.2072K}  and \citet{2023MNRAS.518.3326L} calculated the evolution of the dust particle size distribution, considering collisional fragmentation in a collapsing cloud core, and showed the importance of dust fragmentation for realistically estimating non-ideal MHD effects.

We need to evaluate the relative velocity between dust particles and the critical particle fragmentation velocity in order to determine the evolution of the dust particle size distribution. 
In past studies of dust growth in molecular cloud cores,  gas turbulence was usually considered to be the origin of the relative velocity between dust particles \citep{2009A&A...502..845O,2013MNRAS.434L..70H}.
The critical fragmentation velocity corresponds to the velocity required for most of dust particles to fragment after a collision. 

In Paper I, the turbulence velocity was assumed to be comparable to the speed of sound.
In addition, the critical fragmentation velocity was uniquely determined considering the dust composition and the initial dust particle size distribution.
However, there exist uncertainties in both the turbulent and fragmentation velocities. 

The turbulence velocity or intensity during the core-collapse phase has not been well investigated  
because it is difficult to perform high-resolution numerical simulations and to observe small-scale velocity fluctuations, which are required to understand the properties of turbulence.
In addition, turbulence intensity should  differ among cores.
The fragmentation velocity has been investigated in both laboratory  \citep{1993Icar..106..151B, 2005Icar..178..253W, 2008ApJ...675..764L} and numerical \citep{1997ApJ...480..647D,2013A&A...559A..62W,2021ApJ...915...22H} experiments.
These studies have shown that the fragmentation velocity depends on many factors such as the dust composition and structure (e.g., porosity).
In this study, we use the turbulence intensity and fragmentation velocity as parameters and investigate their influence on the evolution of the dust particle size distribution and non-ideal MHD effects during the core-collapse phase.

This paper is structured as follows.
We describe the equations and the model for calculating the dust particle size distribution in Section~\ref{sec:method}.
The results are presented in Section~\ref{sec:results}.
The implications for the star and disk formation processes and caveats are discussed in Section~\ref{se:discussion}. 
A summary is presented  in Section~\ref{sec:summary}.

\section{Methods}
\label{sec:method}
The methods for calculating the dust particle size evolution and magnetic diffusion coefficients used in this study are similar to those in Paper I. 
Thus, we explain them simply in the following subsections. 

\subsection{Dust coagulation and fragmentation equation}
The evolution of the dust particle size distribution is described by an integro-differential equation involving the dust mass density $\rho_{d} \left(m, t\right)$ and the dust mass $m$ at time $t$.
The dust mass between $m$ and $m+dm$ is expressed by $\rho_{d} \left(m, t\right)\mathrm{d}m$. 
$\rho_{d} \left(m, t\right)$ evolves due to collisional coagulation and fragmentation of dust particles
and this time evolution considering both coagulation and fragmentation  
can be expressed as \citep{1916ZPhy...17..557S,2018MNRAS.475..167B} 
\begin{align}
  \frac{\mathrm{d}\rho_{d}\left(m,t\right)}{\mathrm{d}t}
  & = \frac{1}{2}\int_{0}^{m} m K\left(m-m_{1}, m_{1}\right) n_{d}\left(m-m_{1}, t\right) \notag \\ 
  & \ \ \ \ \times n_{d}\left(m_{1}, t\right) \mathrm{d}m_{1}\notag \\
  & - \int_{0}^{\infty}mK\left(m,m_{1}\right) n_{d}\left(m, t\right) n_{d}\left(m_{1}, t\right) \mathrm{d}m_{1} \notag \\
  & + \frac{1}{2}\int \int_{0}^{\infty} m F\left(m_{1}, m_{2}\right) n_{d}\left(m_{1}, t\right) n_{d} \left(m_{1}, t\right) \notag \\ 
  & \ \ \ \ \times \varphi_{\rm f}\left(m ; m_{1}, m_{2}\right) 
  \mathrm{d}m_{1}\mathrm{d}m_{2} \notag \\ 
  & - \int_{0}^{\infty} m F\left(m,m_{1}\right) n_{d}\left(m, t\right) n_{d}\left(m_{1}, t\right) \mathrm{d}m_{1} \notag \\
  & + \frac{\rho\left(m, t\right)}{\rho_{g}}\frac{\mathrm{d}\rho_{g}}{\mathrm{d} t},
  \label{eq:coag-frag-equation}
\end{align} 
where $n_{d}\left(m, t\right)$ is the number density of dust mass with mass $m$,  $K\left(m, m_{1}\right)$ and $F\left(m, m_{1}\right)$ are the coagulation and fragmentation kernels,
$\varphi_{f} \left(m; m_{1}, m_{2}\right)$ is the distribution function of fragments after a collision between dust particles with masses $m_{1}$ and $m_{2}$,  and $\rho_{g}$ is the gas mass density.
We use the same equations in Paper I for the coagulation kernel $K$ and fragmentation kernel $F$.
The distribution function for fragments $\varphi_{\mathrm{f}}$ is described in Section~\ref{sec:fragmentation_model}.

The last term in equation~(\ref{eq:coag-frag-equation}) represents the change in $\rho_{d} \left(m, t\right)$ due to 
the change in ambient gas density \citep{2019MNRAS.482.2555H} and corresponds to the density evolution in the collapsing core.
We use a one-zone model \citep{2005ApJ...626..627O} to calculate the density evolution of collapsing cores in the same manner as in Paper I.
The gas temperature is also derived using the same barotropic equation as in Paper I.

We assume that each dust particle is compact and spherical.
Although actual dust particles may be distorted or have porosity, we ignore these dust properties.
Thus, the mass of a dust particle with radius $a$ can be expressed as $m = (4\pi/3)\rho_{s}a^{3}$, where $\rho_{s}$ is the internal dust particle density (see \S\ref{sec:parameters}).

\subsection{Relative velocity}
\label{sec:rel_vel} 
Dust particle motion is an important factor in determining dust particle size evolution.
The relative velocity between dust particles is necessary for estimating the collision rate between particles. 
In addition, the relative velocity determines whether dust particles fragment after a collision. 
In this study, we consider turbulence and thermal motion (or Brownian motion) as the origin of the relative velocity between dust particles, as in Paper I.

We assume that dust particles and gas are coupled to each other by collisions and are in thermal equilibrium, i.e.,\ they have the same temperature. 
The relative velocity due to thermal motion  (Brownian motion) between two dust particles with mass $m_{i}$ and $m_{j}$ 
is described by 
\begin{equation}
  \Delta V^{\mathrm{B}}_{ij} = \sqrt{\frac{8 k_{B} T (m_{i} + m_{j}) }{\pi m_{i} m_{j}}},
\end{equation}
where $k_{B}$ is the Boltzmann constant and $T$ is the gas temperature.

Next, we explain our turbulence model.
We define the turnover time for the largest turbulence eddies corresponding to the Jeans length $L_{\mathrm{J}}$ as the sound crossing time,
\begin{equation}
  \tau_{\mathrm{L}} = \frac{L_{\mathrm{J}}}{c_{s}} = \frac{1}{2}\sqrt{\frac{\pi}{G\rho_{g}}},
\end{equation}
where $c_{s}$ is the sound speed. 
In Paper I, the fluctuation velocity of turbulence was assumed to be equal to the speed of sound on any scale.
Although there should be non-negligible velocity fluctuations (or turbulence) in a collapsing core, the turbulence intensity is uncertain. 
Thus, in this study, we parameterize the turbulence intensity to investigate the effect of  turbulence on the evolution of the dust particle size distribution.
The turbulent fluctuation velocity is given by
\begin{equation}
  V_{\mathrm{turb}} = \alpha c_{s}.
\end{equation}
We vary the turbulence intensity using the parameter $\alpha$, which ranges from zero to unity  $(0 \le \alpha \le 1)$.
Since there are uncertainties in turbulence intensity, supersonic turbulence could be realized in a collapsing cloud core. 
However, we expect that supersonic turbulence could dissipate in a short timescale.  
Thus, in this study, we only consider sub-sonic turbulence with $\alpha\le1$.

Assuming a Kolmogorov turbulent cascade, the eddy turnover time on the viscous dissipation scale is described as 
\begin{equation}
  \tau_{\eta} = \frac{\tau_{\mathrm{L}}}{\sqrt{\mathrm{Re}}},
\end{equation}
where $\mathrm{Re} = \nu_{\mathrm{turb}}/\nu_\mathrm{mol}$ is the Reynolds number.
The turbulent viscosity is given by $\nu_{\mathrm{turb}} = V_{\mathrm{turb}} L_{\mathrm{J}} = \alpha c_{s} L_{\mathrm{J}}$. 
The molecular viscosity is described by $\nu_{\mathrm{mol}} =  V_{\mathrm{th}}\, l_{\mathrm{mfp}} /2$, where 
$V_{\mathrm{th}} = \sqrt{8/\pi}c_{s}$ and $l_{\mathrm{mfp}} = m_{\mu}/(\rho_{g} \sigma_{\mathrm{H_{2}}})$ are 
the thermal velocity and the mean free path of gas particles, respectively, $m_{\mu}$ is the mass of a gas molecule, and $\sigma_{\mathrm{H_{2}}} = 2\times 10^{-15} \ \mathrm{cm^{2}}$
is the collisional cross section for $\mathrm{H_{2}}$ gas particles.

The dust dynamics in a turbulent flow is controlled by the stopping time $t_{s}$.
We use the following formula for the stopping time, 
\begin{equation}
  t_{s} = \left\{
    \begin{aligned}
      &\frac{\rho_{s} a}{\rho_{g} v_{\mathrm{th}}}, \ \ \ \ \ \ \  (a < \frac{9}{4}l_{\mathrm{mfp}}) \\
      &\frac{4a^{2}}{9 l_{\mathrm{mfp}} v_{\mathrm{th}} \rho_{g}}. \ \  (a > \frac{9}{4}l_{\mathrm{mfp}}) 
    \end{aligned}
  \right.
\end{equation}
The former is Epstein's law and the latter is Stokes's law.

Using the dust stopping time $t_{s}$ and the eddy turnover time $\tau_{L}$, we define the Stokes number as $St = t_{s} / \tau_{L}$. 
When considering two dust particles with different Stokes numbers $St_{1} = \tau_{1}/\tau_{\mathrm{L}}$ and $St_{2} = \tau_{2}/\tau_{\mathrm{L}}$ $(St_{1} \ge St_{2})$,
the relative velocity between these particles due to  turbulence is given by the following equations \citep{2007A&A...466..413O},
\begin{equation}
  \Delta V_{\mathrm{turb}} = \sqrt{\Delta V^{2}_{\mathrm{I}} + \Delta V^{2}_{\mathrm{II}}},
\end{equation}
\begin{align}
  \Delta V^{2}_{\mathrm{I}}
  &=  \sqrt{\frac{3}{2}} v_{\mathrm{turb}} \frac{St_{1} - St_{2}}{St_{1} + St_{2}} \notag \\
    & \times \left( \frac{St^{2}_{1}}{St^{\ast}_{1,2} + St_{1}} - \frac{St^{2}_{1}}{1 + St_{1}}- \frac{St^{2}_{2}}{St^{\ast}_{1,2} + St_{2}} + \frac{St^{2}_{2}}{1 + St_{2}} \right),
\end{align}
and
\begin{align}
  \Delta V^{2}_{\mathrm{II}}
  & = \sqrt{\frac{3}{2}} v_{\mathrm{turb}} \left( St^{\ast}_{1,2} - St_{\mathrm{min}} \right) \notag \\ 
  & \times \Biggl\{ \frac{ \left(St^{\ast}_{1,2} + St_{\mathrm{min}} \right)  St_{1} + St^{\ast}_{1,2}St_{\mathrm{min} }}
  { \left(St_{1} + St^{\ast}_{1,2}\right) \left(St_{1} + St_{\mathrm{min}}\right) } \notag \\ 
  & + \frac{ \left(St^{\ast}_{1,2} + St_{\mathrm{min}} \right)  St_{2} + St^{\ast}_{1,2}St_{\mathrm{min} }}
  { \left(St_{2} + St^{\ast}_{1,2}\right) \left(St_{2} + St_{\mathrm{min}}\right) } \Biggr\},
\end{align}
where $St_{\mathrm{min}} = \tau_{\eta}/\tau_{\mathrm{L}} = 1/\sqrt{\mathrm{Re}}$ and  
$St^{\ast}_{1,2} = \mathrm{max}\left( St_{\mathrm{min}}, \mathrm{min}\left(1.6 St_{1}, 1\right) \right)$.

Based on this derivation, the relative velocity including both turbulence and Brownian motion is described  by
\begin{equation}
  \label{eq:rel_vel_total}
  \Delta V_{ij} = \sqrt{ \left(\Delta V^{\mathrm{T}}_{ij} \right)^{2} + \left( \Delta V^{\mathrm{B}}_{ij} \right)^{2} }.
\end{equation}

\subsection{Fragmentation model}
\label{sec:fragmentation_model}
Our fragmentation model is constructed based on numerical dust collision calculations \citep{2013A&A...559A..62W} and is essentially the same as in Paper I. 
The distribution function for fragments $\varphi_{\mathrm{f}}$ is the same as the formula in Paper I. 

An important physical quantity for calculating the dust particle size evolution is the velocity $v_{\mathrm{col, crit}}$ required for most of the dust particles to fragment after collision, which is the fragmentation velocity.
In Paper I, we uniquely determined the fragmentation velocity based on the composition of dust particles and the initial size distribution.
However, there is still uncertainty in the fragmentation velocity even when the composition and structure of dust particles are determined.
Thus, this study treats $v_{\mathrm{col, crit}}$ as a parameter, unlike Paper I.
We adopt a range of $1 \le v_{\mathrm{col, crit}} \le 180 \ \mathrm{m\, s^{-1}}$. 

\subsection{Ionization degree and non-ideal magnetic diffusion coefficients}
The magnetic diffusion coefficients determine the efficiency of non-ideal MHD effects and are calculated using the same formula as in Paper I. 
We require the number density of charged particles $n_{j}$ to estimate the magnetic diffusion coefficients. 
We calculate the number density of electrons and ions and the average charge number of each dust particle using the method proposed by \citet{2021A&A...649A..50M}. 
The magnetic diffusion coefficients for the Hall effect and ambipolar diffusion depend on the magnetic flux density.
Thus, we adopt the magnetic flux density as a function of gas number density as \citep{2011ApJ...738..180L}
\begin{equation}
  B = 1.43\times 10^{-7} \sqrt{n_{\rm H}}.
\label{eq:bevo}
\end{equation}
Equation~(\ref{eq:bevo}) can be applied for $n_{\mathrm{H}} \lesssim 10^{11} \ \mathrm{cm^{-3}}$, where the magnetic field and gas are well coupled.
On the other hand,  the magnetic field is dissipated due to non-ideal MHD effects in the high-density region $n_{\mathrm{H}} \gtrsim  10^{11} \ \mathrm{cm^{-3}}$ \citep[][and Paper I]{2010MNRAS.408..322K}.
Thus, equation~(\ref{eq:bevo}) may not be entirely correct in the high-density region, and the magnetic flux density and the magnetic diffusion coefficients may be somewhat overestimated. 

\subsection{Parameters and initial conditions}
\label{sec:parameters}
We assume that the internal density of dust particles is $ \rho_{s} = 2.65 \ \rm{g\, cm^{-3}}$ \citep{2009A&A...502..845O}. 
If dust particles are porous, this value will be lower. 
However, since we ignore porosity, as in Paper I, we use the above value even when we consider coagulation growth and collisional fragmentation.
The dust-to-gas mass ratio is set to $f_{\mathrm{dg}} = 0.01$, as described above. 
The dust and gas are assumed to be perfectly coupled; thus, the dust-to-gas mass ratio $f_{\mathrm{dg}}$ does not change during the calculation, 
which is justified by recent three dimensional simulations including dust motion \citep[e.g.,][]{2020A&A...641A.112L, 2022MNRAS.515.6073K}. 

In Paper I, the MRN dust particle size distribution was adopted as the initial size distribution.
On the other hand, in this study, all initial dust particles are set to have the same size. 
The numerical calculations about  the collision of dust particles \citep{2007ApJ...661..320W,2013A&A...559A..62W,2021ApJ...915...22H} investigated the impact of dust (aggregates) composed of the same sized monomer.
The monomer size is one of the key parameters when considering the dust fragmentation process.
Although the actual dust particles may be  composed of monomers of several sizes, the outcomes of a collision between such particles have not been well understood. 
Therefore, to closely or uniquely relate the past numerical simulations to our fragmentation model, we initially set all the dust particles to have the same size (i.e., same-sized monomers).
We adopt two monomer sizes of $a_{0} = 0.1 \ \mathrm{\mu m}$ and $a_{0} = 0.01 \ \mathrm{\mu m}$ in this study \citep{2022A&A...663A..57T}. 

Therefore, we use three parameters, the monomer size $a_{0}$,  turbulence intensity $\alpha$, and fragmentation velocity $v_{\mathrm{col, crit}}$.
The turbulence intensity and fragmentation velocity have ranges of $\alpha = 0$--$1$ and $v_{\mathrm{col, crit}} = 1$--$180 \ \mathrm{m s^{-1}}$, as listed in Table~\ref{tab:parameter}.
Combining these parameters, we prepare 144 models in total.
Each model name is composed of these three parameters. 
For example, model a01-$\alpha$05-v10 has the parameters $a_{0} = 0.1 \ \mu$\,m, $\alpha = 0.5$, and $v_{\mathrm{col, crit}} = 10 \ \mathrm{m\,s^{-1}}$. 

The calculation starts with a gas number density of $n_{\mathrm{H}} = 10^{4} \ \mathrm{cm^{-3}}$.
As the core collapses, the temperature increases and reaches $T \gtrsim 1500 \ \mathrm{K}$ for $n_{\mathrm{H}} > 10^{14} \ \mathrm{cm^{-3}}$.
At this high temperature, the dust particles evaporate. 
Thus, we stop the calculation at $n_{\mathrm{H}} = 10^{14} \ \mathrm{cm^{-3}}$, as in Paper I. 

\begin{table}
    \caption{Parameters used in this study.}
    \begin{tabular}{c|c} \hline
        $a_{0} \ (\mathrm{\mu m})$ &  0.1, 0.01 \\         $\alpha$  & 0, 0.01, 0.1, 0.2, 0.3, 0.5, 0.8, 1 \\ 
        $v_{\mathrm{col, crit}} \ (\mathrm{m s^{-1}})$ & 1, 10, 30, 50, 80, 100, 120, 150, 180 \\ 
        \hline
    \end{tabular}
    \label{tab:parameter}
\end{table}

\section{results}
\label{sec:results}
In this section, we describe several models with a monomer dust particle size of $a_{0} = 0.1 \ \mathrm{\mu m}$ (\S\ref{sec:fixed}--\ref{sec:suma01}) before we show the models with $a_{0} = 0.01 \ \mathrm{\mu m}$ (\S\ref{sec:a001}). 
In \S\ref{sec:fixed} we first describe a special model with $a_{0} = 0.1 \ \mathrm{\mu m}$ (model a01) that does not consider dust particle size evolution. 
We then present the models that do consider dust particle size evolution. 
We refer to model a01-$\alpha$1-v10 as the fiducial model and explain it in \S\ref{sec:fiducial}.  
We then describe the results for each model with $a_{0} = 0.1 \ \mathrm{\mu m}$ in \S\ref{sec:other} and \S\ref{sec:suma01}.
Finally, we comment on the models with $a_{0} = 0.01 \ \mathrm{\mu m}$ in \S\ref{sec:a001}. 

\subsection{Case of fixed dust particle size distribution}
\label{sec:fixed}
%%%%%%
% Fig. 1
%%%%%%
\begin{figure}
  \centering
  \includegraphics[width=80mm]{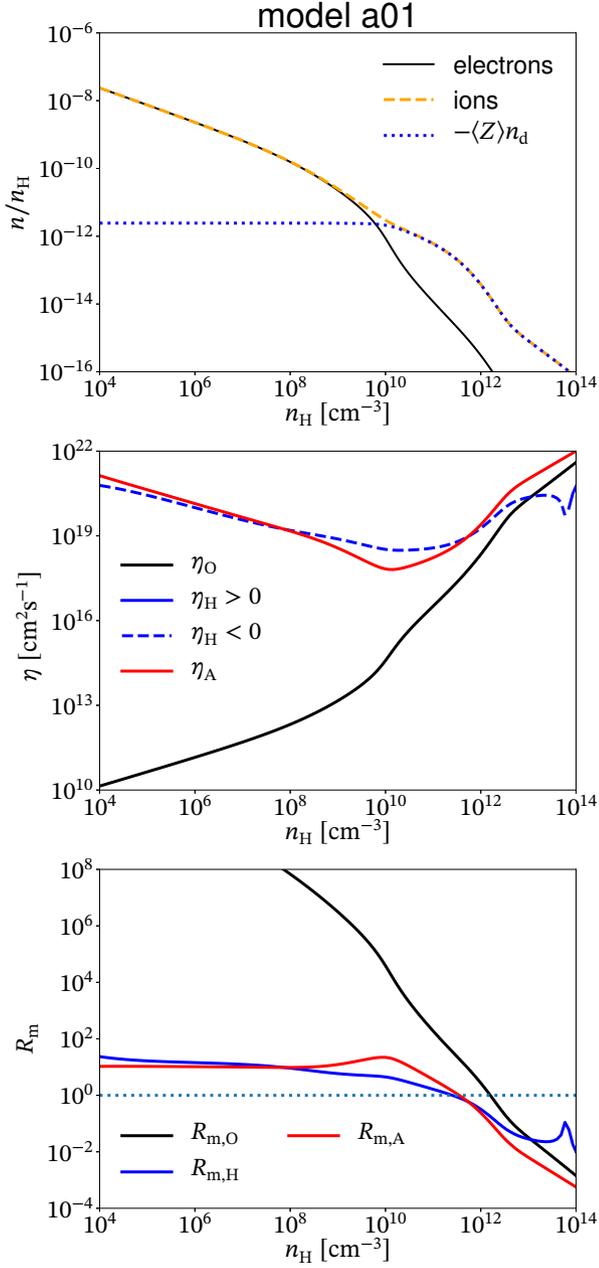}
  \caption{Abundance of charged particles (top), magnetic diffusion coefficients (middle), and magnetic Reynolds number (bottom) against number density for a model that does not consider dust particle size evolution (model a01).}
  \label{fig:a01_abund_eta}
\end{figure}

This subsection describes the results when dust particle size evolution is not considered (model a01). 
The turbulence intensity and fragmentation velocity are not necessary for model a01 because dust particle growth is ignored. 
The dust particle size is fixed at $a = 0.1 \ \mathrm{\mu m}$ for this model. 
The top panel of Figure~\ref{fig:a01_abund_eta} shows the abundance of charged particles (ions and electrons) for model a01.
This panel also shows the net dust charge density $\langle Z \rangle n_{\mathrm{d}}$, where $\langle Z \rangle$ is the average dust charge number and  $n_{\mathrm{d}}$ is the total dust particle number density.
For $n_{\mathrm{H}} \lesssim 10^{10} \ \mathrm{cm^{-3}}$, the electron and ion abundance are almost the same.
On the other hand, for $n_{\mathrm{H}} \gtrsim 10^{10} \ \mathrm{cm^{-3}}$, the abundance of electrons decreases because electron adsorption on the dust particle surface is more efficient than the generation of free electrons by ionization.
In addition, electrons are more selectively adsorbed on the dust particle surface than ions because the thermal velocity of electrons is greater than that of ions. 
Thus, electrons collide more frequently with dust particles than ions. 
Then, dust particles become the main charge carrier instead of electrons for $n_{\mathrm{H}} \gtrsim 10^{10} \ \mathrm{cm^{-3}}$. 

The middle panel in Figure~\ref{fig:a01_abund_eta} shows the magnetic diffusion coefficients for model a01.
Although the Ohmic coefficient $\eta_{O}$ monotonically increases as the density increases, $\eta_{O}$ is always smaller than the ambipolar diffusion coefficient $\eta_{A}$.
For $n_{\mathrm{H}} \lesssim 10^{8} \ \mathrm{cm^{-3}}$ and $n_{\mathrm{H}} \gtrsim 10^{12} \ \mathrm{cm^{-3}}$, $\eta_{A}$ is larger than the other two coefficients.
The Hall coefficient $\eta_{H}$ is the largest among the three diffusion coefficients in the density range $10^{8} \lesssim n_{\mathrm{H}} \lesssim 10^{12} \ \mathrm{cm^{-3}}$.
The sign of $\eta_{H}$ is always negative except above a density of $n_{\mathrm{H}} \simeq 10^{14} \ \mathrm{cm^{-3}}$.

We calculate the magnetic Reynolds number $R_{\mathrm{m}} = VL/\eta$ using the magnetic diffusion coefficients. 
We use the Jeans length $L_{\mathrm{J}}$ and the free-fall velocity $v_{\mathrm{ff}} = \sqrt{4\pi G L_{\mathrm{J}}^{2} \rho_{g}/3}$ as the typical length-scale $L$ and  typical velocity $V$, as in Paper I and \citet{2007ApJ...670.1198M}. 
Note that evaluating the Hall effect in terms of the magnetic Reynolds number is not appropriate because the Hall effect is not a dissipative effect.
However, we note that the Hall effect may have an impact on the formation and evolution of a circumstellar disk \citep{2015ApJ...810L..26T, 2019A&A...631A..66M, 2021MNRAS.507.2354W}.

The bottom panel in Figure~\ref{fig:a01_abund_eta} shows the magnetic Reynolds number.
For $n_{\mathrm{H}} < 10^{12} \ \mathrm{cm^{-3}}$, the magnetic Reynolds number is $R_{\mathrm{m}} \gtrsim 1$--$10$ and non-ideal MHD effects are not significant, indicating little dissipation of the magnetic field.
On the other hand, $R_{\mathrm{m}} \lesssim 1$ is realized for both ambipolar diffusion in the density range $n_{\mathrm{H}} \gtrsim 3.0\times 10^{11} \ \mathrm{cm^{-3}}$
and Ohmic dissipation in the density range $n_{\mathrm{H}} \gtrsim 2.0\times 10^{12} \ \mathrm{cm^{-3}}$.
In other words, non-ideal MHD effects, especially Ohmic diffusion and ambipolar diffusion, can remove the magnetic flux from high-density regions for model a01.

\subsection{Fiducial model}
\label{sec:fiducial}
%%%%%%
% Fig. 2
%%%%%%
\begin{figure}
  \centering
  \includegraphics[width=80mm]{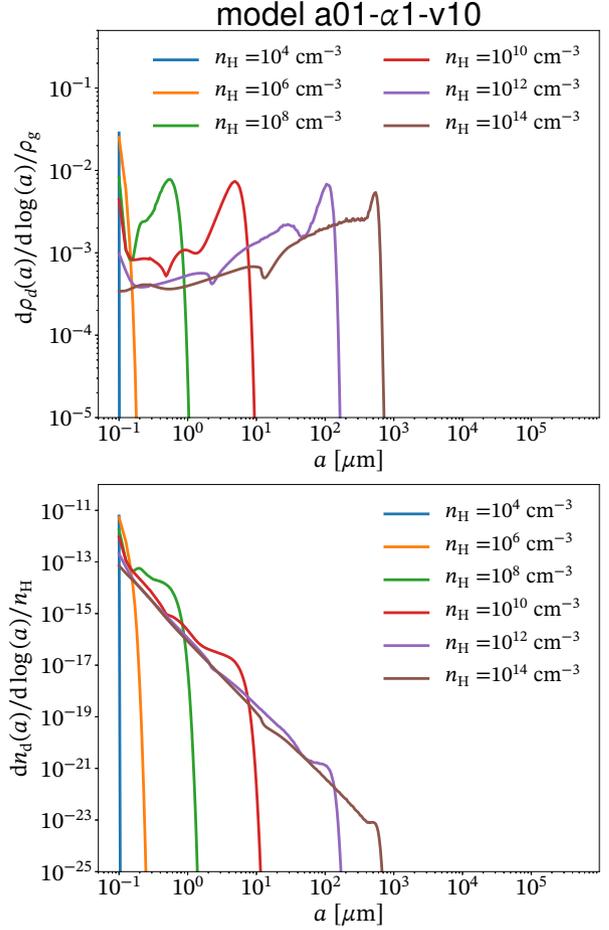}
  \caption{Dust particle size evolution for model a01-$\alpha$1-v10. 
  The dust mass density (top) and dust number density (bottom) are plotted against the dust particle size.  
Each line corresponds to the dust particle size distribution at each epoch. 
}
  \label{fig:a01_alpha1_v10_dist}
\end{figure}

%%%%%%
% Fig. 3
%%%%%%
\begin{figure}
  \centering
  \includegraphics[width=80mm]{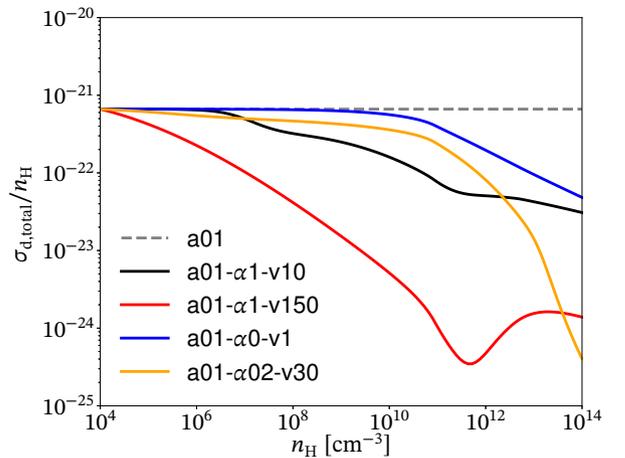}
  \caption{Total dust cross-section $\sigma_{\mathrm{d, total}}$ per hydrogen nucleus (H) against number density for models with a monomer size of $a_{0} = 0.1 \ \rm{\mu m}$.}
  \label{fig:a01_dust_cross_section}
\end{figure}

%%%%%%
% Fig. 4
%%%%%%
\begin{figure}
  \centering
  \includegraphics[width=80mm]{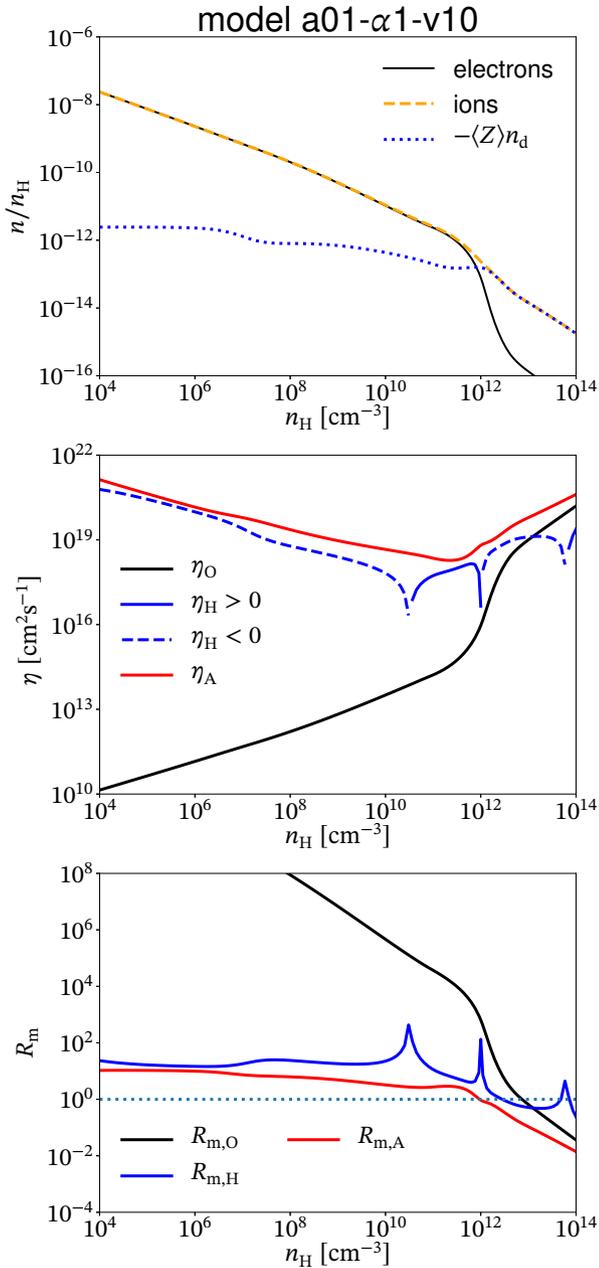}
  \caption{As Fig~\ref{fig:a01_abund_eta} but for model a01-$\alpha$1-v10.
}
  \label{fig:a01_alpha1_v10_abund_eta}
\end{figure}

In this section, we describe the results for model a01-$\alpha$1-v10.
For this model, the turbulence intensity corresponds to the speed of sound ($\alpha = 1$) and the fragmentation velocity is relatively small
($v_{\mathrm{col, crit}} = 10\, \mathrm{m\,s^{-1}}$).

Figure~\ref{fig:a01_alpha1_v10_dist} shows the evolution of the dust particle size distribution for a01-$\alpha$1-v10.
The upper panel shows the dust mass density normalized by the gas mass density 
and the lower panel presents the dust number density normalized by the gas number density.
In the early stage of collapse for $n_{\mathrm{H}} \lesssim 10^{6} \ \mathrm{cm^{-3}}$, little dust particle coagulation occurs.
For $n_{\mathrm{H}} \gtrsim 10^{6} \ \mathrm{cm^{-3}}$, dust particles grow by collisional coagulation and the maximum size of dust particles becomes large as the density increases. 
Small dust particles are also produced by collisional fragmentation.
For $n_{\mathrm{H}} \gtrsim 10^{10} \ \mathrm{cm^{-3}}$,  collisions between dust particles frequently occur due to the high density, which significantly promotes dust growth.
The maximum dust particle size reaches $a \simeq 700 \ \mathrm{\mu m}$ at $n_{\mathrm{H}} = 10^{14} \ \mathrm{cm^{-3}}$.

For the dust mass density, the distribution exhibits a peak at the largest dust particle size  (Fig.~\ref{fig:a01_alpha1_v10_dist} top). 
On the other hand, although the dust number density of the initial dust particle size $a = 0.1 \ \rm{\mu m}$ decreases due to dust growth as the gas density increases,  the dust number density has a peak at the smallest dust particle size ($\sim 0.1 \ \rm{\mu m}$, Fig~\ref{fig:a01_alpha1_v10_dist} bottom) at each epoch.

Figure~\ref{fig:a01_dust_cross_section} shows the evolution of the total (geometrical) dust cross section ({\bf $\sigma_{\mathrm{d, total}} = \int \pi a^{2} n_{d}(a) \mathrm{d}a $}),
in which the total cross sections divided by $n_{\rm{H}}$ are plotted to show the quantity per hydrogen nucleus (H).
The total cross section of the dust is a key factor determining the abundance of charged particles. 
Since the charged particles are absorbed by the dust particles, the degree of ionization decreases and magnetic diffusion becomes stronger as the total cross section increases. 
For $n_{\mathrm{H}} \lesssim 10^{6} \ \mathrm{cm^{-3}}$, the total dust cross section in a01-$\alpha$1-v10 is the same as in the model with a fixed dust particle size (model a01)
because dust coagulation growth hardly occurs.
For $n_{\mathrm{H}} \gtrsim 10^{6} \ \mathrm{cm^{-3}}$, 
the dust cross section decreases because coagulation growth of dust particles is promoted and dust particles size increases as the density increases.

The top panel in Figure~\ref{fig:a01_alpha1_v10_abund_eta} shows the abundance of charged particles for a01-$\alpha$1-v10.
The abundances of electrons and ions are almost the same in the range $n_{\mathrm{H}} \lesssim 10^{12} \ \mathrm{cm^{-3}}$.
For $n_{\mathrm{H}} \gtrsim 10^{12} \ \mathrm{cm^{-3}}$,  adsorption of electrons and ions  onto the dust particle surfaces becomes efficient.
However, electrons and ions are more abundant in a01-$\alpha$1-v10 than in a01 (see Fig.~\ref{fig:a01_abund_eta} top panel). 
This is because, for a01-$\alpha$1-v10,  the abundance or total cross section of dust particles decreases due to coagulation growth 
(Fig.~\ref{fig:a01_alpha1_v10_dist} bottom panel and Fig.~\ref{fig:a01_dust_cross_section}).
Thus,  adsorption of charged particles on the dust particle surfaces is less efficient in a01-$\alpha$1-v10 than a01.

The middle panel in Figure~\ref{fig:a01_alpha1_v10_abund_eta} shows the magnetic diffusion coefficients for a01-$\alpha$1-v10.
The Ohmic coefficient $\eta_{O}$ rapidly increases around $n_{\mathrm{H}} \simeq 10^{12} \ \mathrm{cm^{-3}}$ because the electron abundance sharply decreases.
However, $\eta_{O}$ is always smaller than the ambipolar diffusion coefficient $\eta_{A}$.
$\eta_{A}$ is the largest among all the coefficients over the whole density range.
The Hall coefficient $\eta_{H}$ reverses sign three times in the range $10^{10} < n_{\mathrm{H}} < 10^{14} \ \mathrm{cm^{-3}}$.

The bottom panel in Figure~\ref{fig:a01_alpha1_v10_abund_eta} shows the magnetic Reynolds number $R_{\mathrm{m}}$ for a01-$\alpha$1-v10.
$R_{\mathrm{m}}$ drops below unity in the range $n_{\mathrm{H}} \gtrsim 7.0\times 10^{11} \ \mathrm{cm^{-3}}$ for ambipolar diffusion and in the range $n_{\mathrm{H}} \gtrsim 10^{13} \ \mathrm{cm^{-3}}$ for Ohmic diffusion.
Thus, dissipation of the magnetic field occurs for  $n_{\mathrm{H}} \gtrsim 10^{12} \ \mathrm{cm^{-3}}$. 
The density at which the magnetic Reynolds number reaches $R_{\mathrm{m}} = 1$  is higher for a01-$\alpha$1-v10 than for a01, meaning that magnetic dissipation is more effective in a01 than in a01-$\alpha$1-v10. 

\subsection{Other models}
\label{sec:other}
%%%%%%
% Fig. 5
%%%%%%
\begin{figure}
  \centering
  \includegraphics[width=80mm]{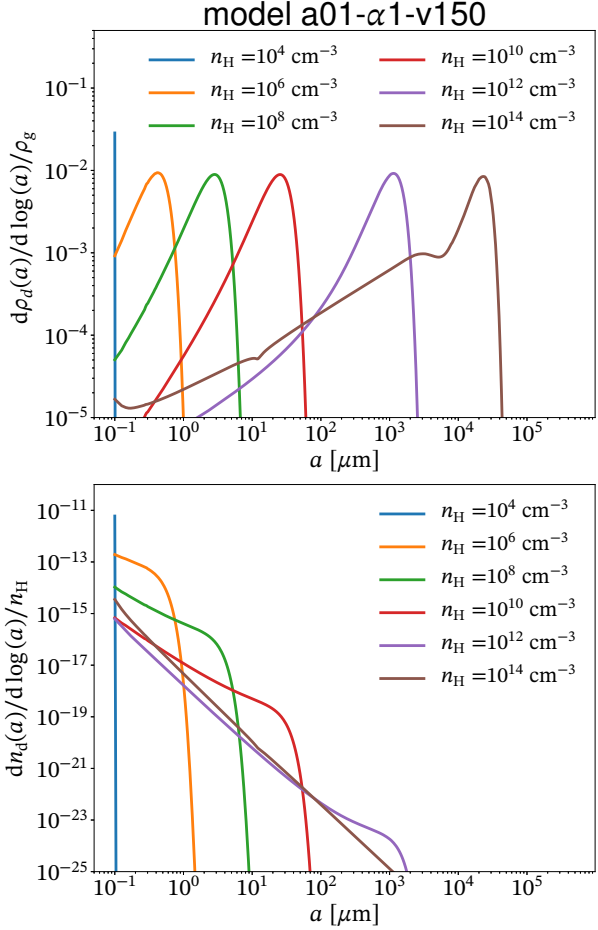}
  \caption{As Fig.~\ref{fig:a01_alpha1_v10_dist} but for a01-$\alpha$1-v150.}
  \label{fig:a01_alpha1_v150_dist}
\end{figure}

%%%%%%
% Fig. 6
%%%%%%
\begin{figure}
  \centering
  \includegraphics[width=80mm]{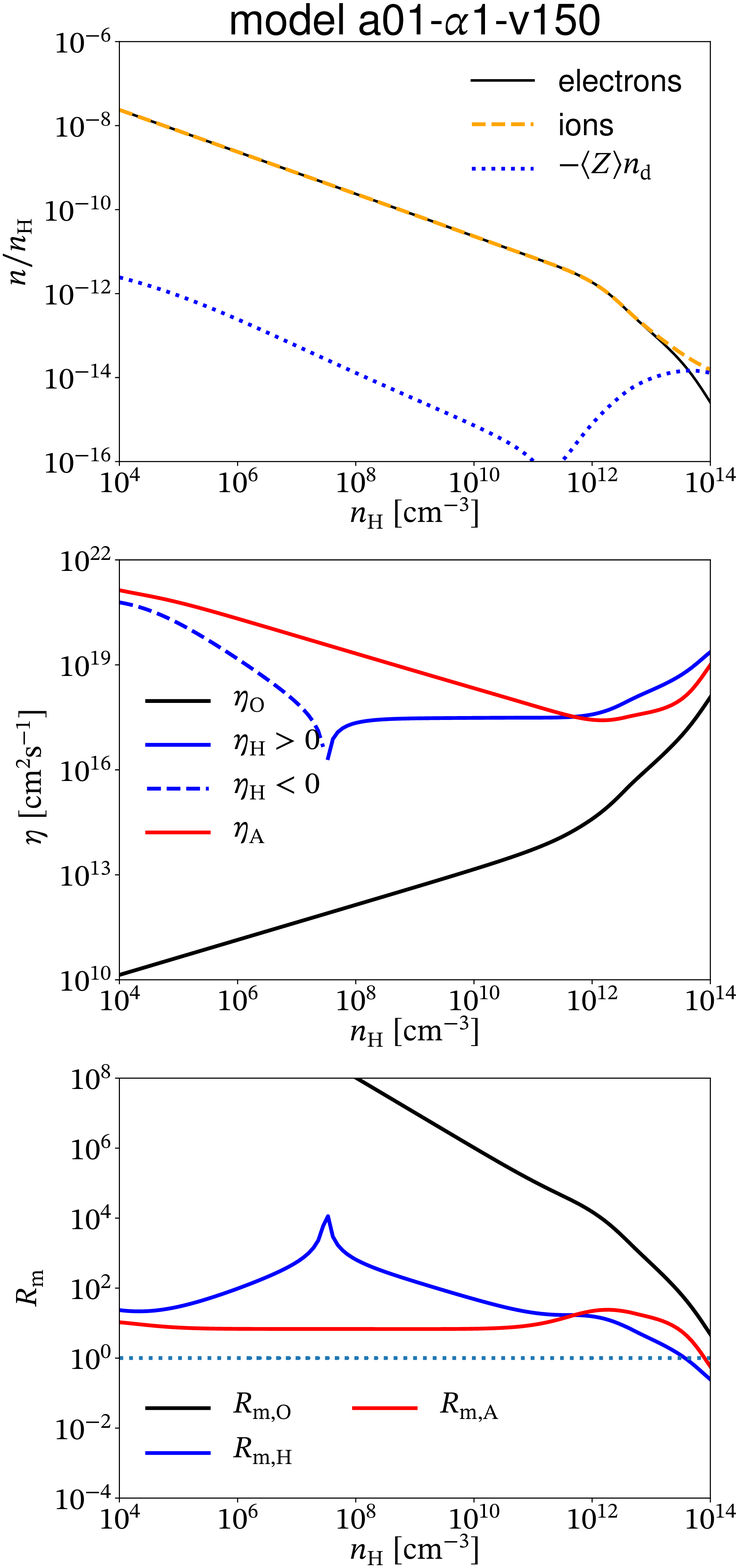}
  \caption{Same as Fig.~\ref{fig:a01_abund_eta} but for a01-$\alpha$1-v150.}
  \label{fig:a01_alpha1_v150_abund_eta}
\end{figure}

In this section, we present some other models with a monomer size of $a_{0}=0.1\,\mu$m. 

\subsubsection{Strong turbulence and large fragmentation velocity case}
First, we describe the results for model a01-$\alpha$1-v150. 
The fragmentation velocity ($v_{\mathrm{col, crit}} = 150 \, \mathrm{m\,s^{-1}}$) for a01-$\alpha$1-v150 is much larger than that for the fiducial model a01-$\alpha$1-v10 ($v_{\mathrm{col, crit}} = 10 \, \mathrm{m\,s^{-1}}$). 
On the other hand, the turbulence intensity ($\alpha=1$) is the same for both models. 
Thus, collisional fragmentation is less likely to occur in a01-$\alpha$1-v150 due to the high velocity required for fragmentation.
Figure~\ref{fig:a01_alpha1_v150_dist} indicates rapid dust growth compared to the fiducial model (Fig.~\ref{fig:a01_alpha1_v10_dist}).
The maximum or peak dust particle size at any epoch is larger for a01-$\alpha$1-v150 than for a01-$\alpha$1-v10
(top panels in Figs~\ref{fig:a01_alpha1_v10_dist} and \ref{fig:a01_alpha1_v150_dist}).
On the other hand, the dust number density for a01-$\alpha$1-v150 is smaller than for a01-$\alpha$1-v10 at any epoch (bottom  panels of Figs~\ref{fig:a01_alpha1_v10_dist} and \ref{fig:a01_alpha1_v150_dist}).
We confirmed that, for a01-$\alpha$1-v150, collisional fragmentation rarely occurs up to $n_{\mathrm{H}} \simeq 10^{12} \ \mathrm{cm^{-3}}$.
On the other hand, for $n_{\mathrm{H}} \gtrsim 10^{12} \ \mathrm{cm^{-3}}$, the dust mass (Fig.~\ref{fig:a01_alpha1_v150_dist} top) and number density (Fig.~\ref{fig:a01_alpha1_v150_dist} bottom) in the range $a\lesssim10$--$100\,\mathrm{\mu m}$  increase due to fragmentation.

The total dust cross section in a01-$\alpha$1-v150 continues to decrease from the early stages until $n_{\mathrm{H}} \simeq 10^{12} \ \mathrm{cm^{-3}}$ (Fig~\ref{fig:a01_dust_cross_section}).
In the range $10^{12} \lesssim n_{\mathrm{H}} \lesssim 10^{13} \ \mathrm{cm^{-3}}$, the total dust cross section increases due to fragmentation. 
For $n_{\mathrm{H}} \gtrsim 10^{13} \ \mathrm{cm^{-3}}$, the total cross section slightly decreases, which is determined by a subtle balance between  coagulation growth and collisional fragmentation.

The top panel in Figure~\ref{fig:a01_alpha1_v150_abund_eta} shows the abundance of charged particles for a01-$\alpha$1-v150.
The electron abundance is almost the same as the ion abundance for  $n_{\mathrm{H}} \lesssim 10^{13}\, \mathrm{cm^{-3}}$.
The density at which a difference in abundance between electrons and ions appears is higher for a01-$\alpha$1-v150 
than for the fiducial model a01-$\alpha$1-v10 (Fig.~\ref{fig:a01_alpha1_v10_abund_eta}) because the total cross section of dust particles for a01-$\alpha$1-v150 significantly decreases due to rapid dust particle growth (Fig.~\ref{fig:a01_dust_cross_section}).
As a result, electron adsorption on dust particle surfaces is less efficient for a01-$\alpha$1-v150 than for the fiducial model. 
On the other hand, for $n_{\mathrm{H}} \gtrsim 10^{12} \ \mathrm{cm^{-3}}$, collisional fragmentation between large dust particles produces small dust particles (Fig.~\ref{fig:a01_alpha1_v150_dist}) and increases the total dust cross section (Fig.~\ref{fig:a01_dust_cross_section}). 
As a result, the abundance of electrons rapidly decreases in the range $n_{\mathrm{H}} \gtrsim 10^{13} \ \mathrm{cm^{-3}}$.

The middle panel in Figure~\ref{fig:a01_alpha1_v150_abund_eta} shows the magnetic diffusion coefficients for a01-$\alpha$1-v150.
For $n_{\mathrm{H}} \lesssim 10^{12} \ \mathrm{cm^{-3}}$, the ambipolar diffusion coefficient $\eta_{A}$ is the largest among the three coefficients.
For $n_{\mathrm{H}} \lesssim 10^{10} \ \mathrm{cm^{-3}}$, $\eta_{A}$ for a01-$\alpha$1-v150 is not significantly different from  that for a01 and a01-$\alpha$1-v10 (fiducial model). 
On the other hand, for $n_{\mathrm{H}} \gtrsim 10^{10} \ \mathrm{cm^{-3}}$, $\eta_{A}$ is smaller for a01-$\alpha$1-v150 than for a01 and a01-$\alpha$1-v10.
In the range $n_{\mathrm{H}} \gtrsim 10^{12} \ \mathrm{cm^{-3}}$, $\eta_{H}$ becomes largest.
The sign of $\eta_{H}$ for a01-$\alpha$1-v150 changes from negative to positive at lower density than for a01 and a01-$\alpha$1-v10. 

The bottom panel in Figure~\ref{fig:a01_alpha1_v150_abund_eta} shows that the magnetic Reynolds number for a01-$\alpha$1-v150 is always above unity except around $n_{\mathrm{H}} \simeq 10^{14} \ \mathrm{cm^{-3}}$.
Thus, when  the fragmentation velocity $v_{\mathrm{col, crit}}$ is large, magnetic dissipation due to non-ideal MHD effects is marginally effective only in the higher-density region of $n_{\mathrm{H}} \gtrsim 10^{14} \ \mathrm{cm^{-3}}$.

%%%%%%
% Fig. 7
%%%%%%
\begin{figure}
  \centering
  \includegraphics[width=80mm]{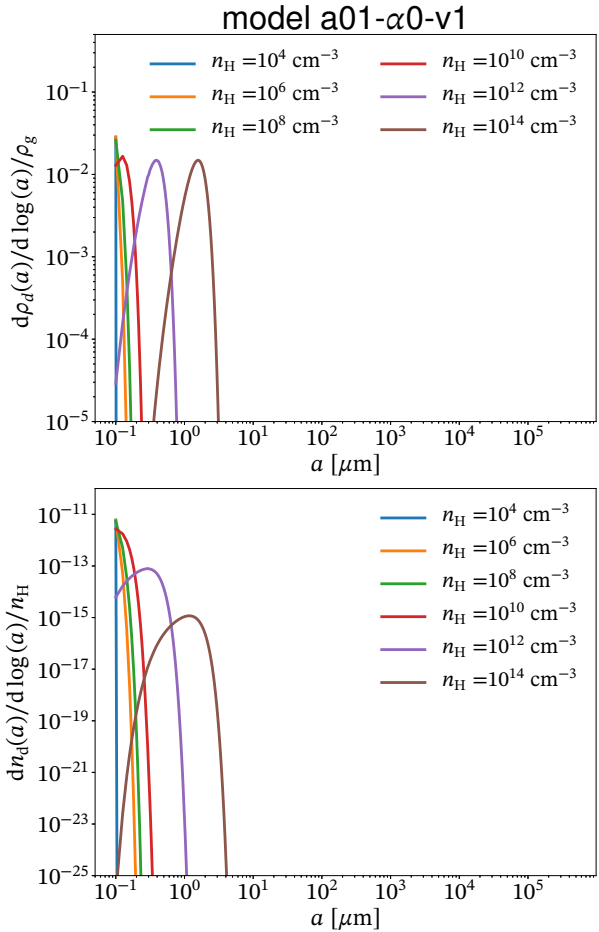}
  \caption{As Fig.~\ref{fig:a01_alpha1_v10_dist} but for a01-$\alpha$0-v1.}
  \label{fig:a01_alpha0_v_dist}
\end{figure}

%%%%%%
% Fig. 8
%%%%%%
\begin{figure}
  \centering
  \includegraphics[width=80mm]{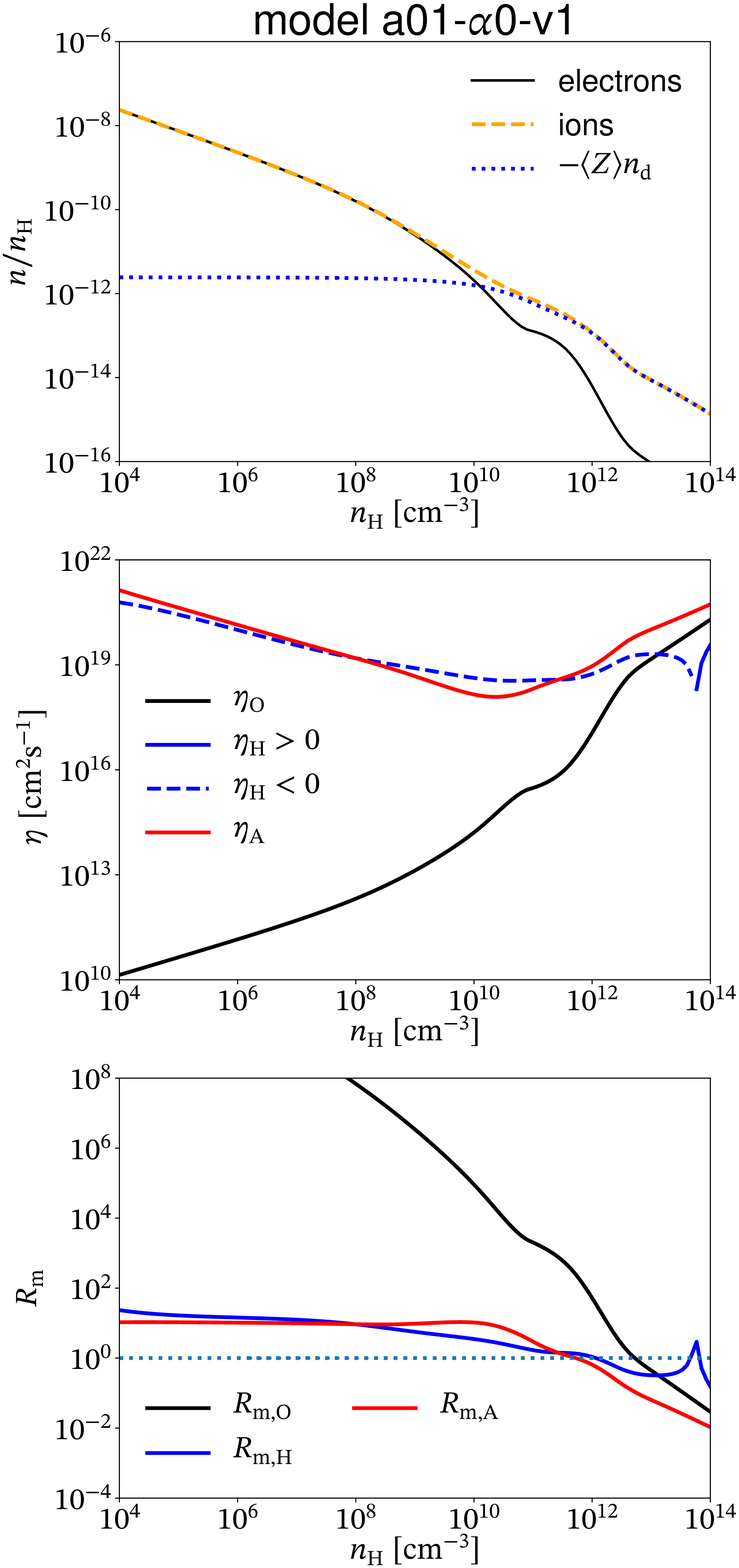}
  \caption{Same as Fig.~\ref{fig:a01_abund_eta} but for a01-$\alpha$0-v1.}
  \label{fig:a01_alpha0_v1_abund_eta}
\end{figure}

\subsubsection{No turbulence case}
\label{sec:a01a0v1}
Next, we present the results for model a01-$\alpha$0-v1.
bf This model has no turbulence  $\alpha=0$.
Thus, only Brownian motion contributes to the relative velocity of the dust particles.
Figure~\ref{fig:a01_alpha0_v_dist} shows the dust particle size evolution for a01-$\alpha$0-v1.
In this model, there is no significant growth of dust particles in the range $n_{\mathrm{H}} \lesssim 10^{10}\, \mathrm{cm^{-3}}$.
For $n_{\mathrm{H}} \gtrsim 10^{10}\, \mathrm{cm^{-3}}$, the collision frequency between dust particles becomes high due to the high density, and dust particle growth is promoted.
However, the maximum size of dust particles is not as large as in the case when turbulence is included.
The maximum size of  dust particles for a01-$\alpha$0-v1 is as small as $a \sim 30\, \mathrm{\mu m}$ at $n_{\mathrm{H}} = 10^{14}\, \mathrm{cm^{-3}}$.
On the other hand, dust particles with sizes of $a \simeq 0.1 \rm{\mu m}$ are depleted compared to the fiducial model a01-$\alpha$1-v10 
because fragmentation does not occur in the absence of turbulence.
In addition, unlike the model with turbulence, the peak size of the dust mass density distribution (Fig.~\ref{fig:a01_alpha0_v_dist} top) is almost the same as that of the dust number density distribution (Fig.~\ref{fig:a01_alpha0_v_dist} bottom) at any epoch.
The dust particle size evolution in the absence of turbulence found in this study is consistent with that in \citet{2009MNRAS.399.1795H}.

The evolution of the total dust cross section for a01-$\alpha$0-v1 is almost the same as for the model with a fixed dust particle size (model a01) up to $n_{\mathrm{H}} \simeq 10^{10}\, \mathrm{cm^{-3}}$
(Fig.~\ref{fig:a01_dust_cross_section}).
Although for $n_{\mathrm{H}} \gtrsim 10^{10}\, \mathrm{cm^{-3}}$, the total cross section for a01-$\alpha$0-v1 decreases due to the increased frequency of dust collisions in the high-density region, it is always larger than for the fiducial model a01-$\alpha$1-v10.

The top panel of Figure~\ref{fig:a01_alpha0_v1_abund_eta} shows the abundance of charged particles for a01-$\alpha$0-v1.
For $n_{\mathrm{H}} \lesssim 10^{10} \ \mathrm{cm^{-3}}$, the electron abundance is almost the same as the ion abundance. 
The density at which the adsorption of electrons on the dust particle surfaces become efficient is almost the same as in the case without dust particle growth (a01).
This is because the total dust cross section does not significantly change for $n_{\mathrm{H}} \lesssim 10^{10} \ \mathrm{cm^{-3}}$ (Fig.~\ref{fig:a01_dust_cross_section}).
However, for $n_{\mathrm{H}} \gtrsim 10^{10}\, \mathrm{cm^{-3}}$, electrons are more abundant for a01-$\alpha$0-v1 than for a01 because the dust particles grow slightly and the abundance or the total cross section of dust particles decreases. 

The middle panel in Figure~\ref{fig:a01_alpha0_v1_abund_eta} shows that the magnetic diffusion coefficients are smaller for a01-$\alpha$0-v1 than for a01 in the range $n_{\mathrm{H}} \gtrsim 10^{11} \ \mathrm{cm^{-3}}$  due to the difference in electron abundance. 
Thus, as shown in the bottom panel in Figure~\ref{fig:a01_alpha0_v1_abund_eta},  the density at which the magnetic Reynolds number is below unity is slightly higher for a01-$\alpha$0-v1 than for a01.
Although there is a slight difference between the model without turbulence and that without dust particle growth, for both models, non-ideal MHD effects could act to remove the magnetic field in high-density regions.

%%%%%%
% Fig. 9
%%%%%%
\begin{figure}
  \centering
  \includegraphics[width=80mm]{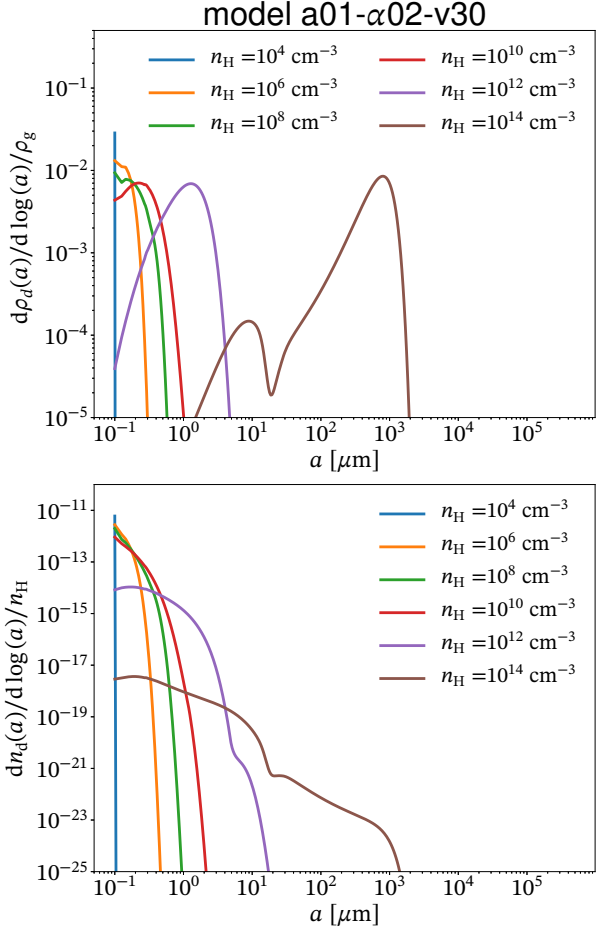}
  \caption{Same as Fig.~\ref{fig:a01_alpha1_v10_dist} but for a01-$\alpha$02-v30.}
  \label{fig:a01_alpha02_v30_dist}
\end{figure}

%%%%%%
% Fig. 10
%%%%%%
\begin{figure}
  \centering
  \includegraphics[width=80mm]{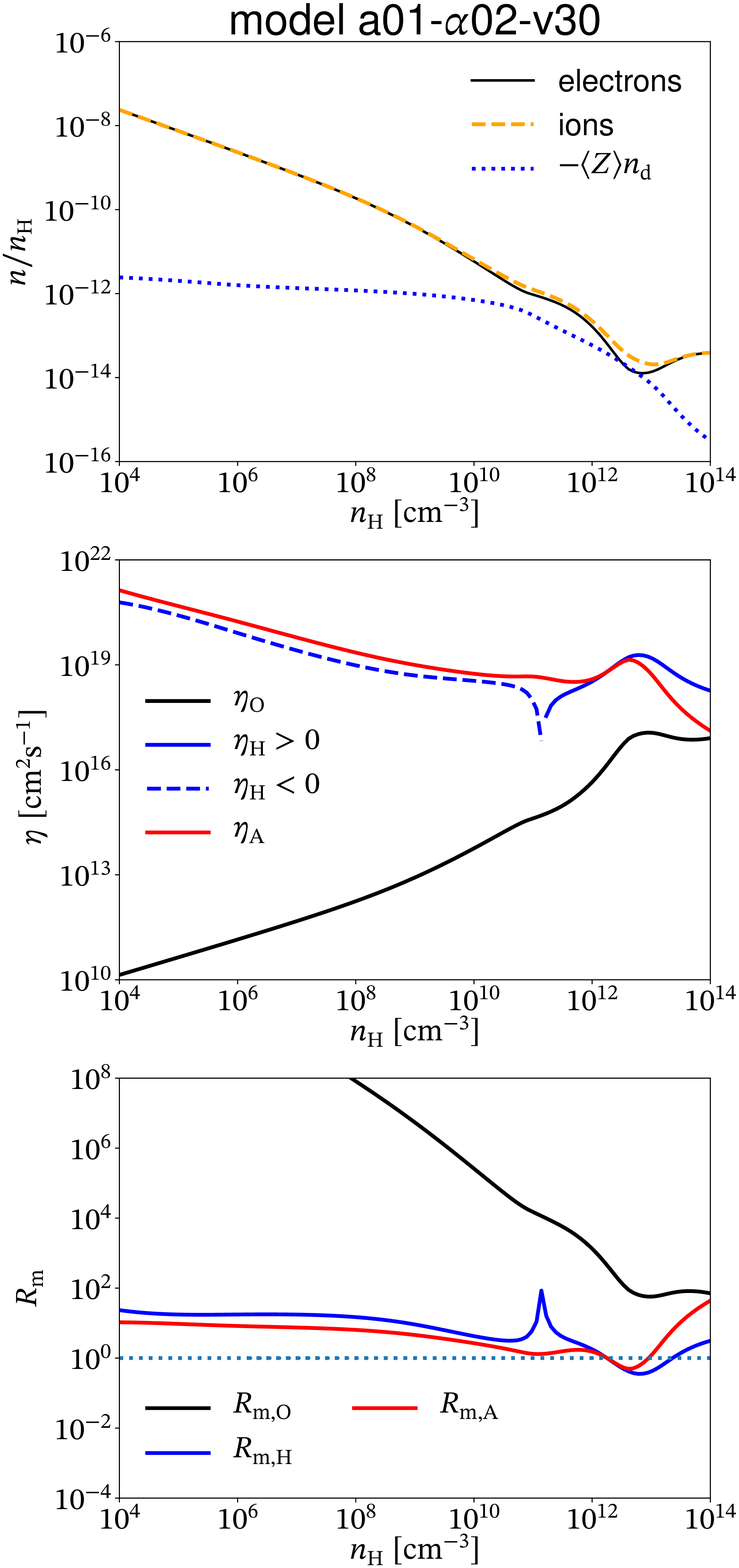}
  \caption{As Fig.~\ref{fig:a01_abund_eta} but for a01-$\alpha$02-v30.}
  \label{fig:a01_alpha02_v30_abund_eta}
\end{figure}

\subsubsection{Other characteristic model}
\label{sec:a01a02v30}
Finally, we describe the results for model a01-$\alpha$02-v30. 
The turbulence intensity is $\alpha=0.2$ and the fragmentation velocity is $v_{\mathrm{col, crit}}=30\, \mathrm{m\,s^{-1}}$ for this model. 
The behavior of the magnetic Reynolds number in this model is considerably different from other models in the high-density region.   
The dust particle size evolution for this model has a different impact on the non-ideal MHD effects than for the models described above.
Figure~\ref{fig:a01_alpha02_v30_dist} plots the dust particle size evolution for a01-$\alpha$02-v30, 
and indicates that dust particles do not grow significantly for $n_{\mathrm{H}} \lesssim 10^{10}\, \mathrm{cm^{-3}}$.
In addition, the maximum size of dust particles is as small as $a \simeq 1.0 \ \mathrm{\mu m}$ at $n_{\mathrm{H}} = 10^{10} \ \mathrm{cm^{-3}}$.
For $n_{\mathrm{H}} \gtrsim 10^{10} \ \mathrm{cm^{-3}}$, collisional growth of dust particles proceeds. 
The dust particles grow rapidly without fragmentation especially in the range $ 10^{12}  \ \mathrm{cm^{-3}} \lesssim n_{\mathrm{H}} \lesssim 10^{14} \ \mathrm{cm^{-3}}$. 
Then, the dust number density significantly decreases because collisions between dust particles are frequent in such a high-density region.
The dust particles grow to about $a \simeq 2\times 10^{3}\, \mathrm{\mu m}$ at $n_{\mathrm{H}} = 10^{14}\, \mathrm{cm^{-3}}$.

For $n_{\mathrm{H}} \lesssim 10^{11} \ \mathrm{cm^{-3}}$, the total dust cross section slowly decreases (Fig.~\ref{fig:a01_dust_cross_section}).
Then, for $n_{\mathrm{H}} \gtrsim 10^{12} \ \mathrm{cm^{-3}}$, the total dust cross section is significantly reduced due to the rapid growth of dust particles.

The top panel in Figure~\ref{fig:a01_alpha02_v30_abund_eta} indicates that the electron abundance is not significantly reduced 
in the range $n_{\mathrm{H}} \gtrsim 10^{12} \, \mathrm{cm^{-3}}$, unlike for the fiducial model (Fig.~\ref{fig:a01_alpha1_v10_abund_eta}). 
For $n_{\mathrm{H}} \gtrsim 10^{13}\, \mathrm{cm^{-3}}$, the electron and ion abundances increase because the dust abundance and the total cross section decrease due to the rapid collisional growth of dust particles.

The middle panel in Figure~\ref{fig:a01_alpha02_v30_abund_eta} shows that $\eta_{A}$ is largest among the three magnetic diffusion coefficients for $n_{\mathrm{H}} \lesssim 10^{12} \ \mathrm{cm^{-3}}$,  while $\eta_{H}$ is the largest for $n_{\mathrm{H}} \gtrsim 10^{12} \ \mathrm{cm^{-3}}$.
For $n_{\mathrm{H}} \gtrsim 10^{13} \ \mathrm{cm^{-3}}$, all the magnetic diffusion coefficients decrease due to the increase in electron and ion abundances. 
The bottom panel in Figure~\ref{fig:a01_alpha02_v30_abund_eta} shows that the magnetic Reynolds number for ambipolar diffusion $R_{\mathrm{m, A}}$ is marginally below unity in the range $10^{12} \lesssim n_{\mathrm{H}} \lesssim 10^{13} \ \mathrm{cm^{-3}}$.
However, $R_{\mathrm{m, A}}$ is again above unity for $n_{\mathrm{H}} \gtrsim 10^{13} \ \mathrm{cm^{-3}}$.
This implies that removal of the magnetic field in the high-density region is not very effective due to the rapid growth of dust particles.
Note that, unlike this model  (a01-$\alpha$02-v30), the magnetic Reynolds number continues to decrease after it becomes below unity in the models described above.

\subsection{Short summary of monomer size  $a_{0} = 0.1 \ \mathrm{\mu m}$}
\label{sec:suma01}
%%%%%%
% Fig. 11
%%%%%%
\begin{figure*}
  \centering
  \includegraphics[width=160mm]{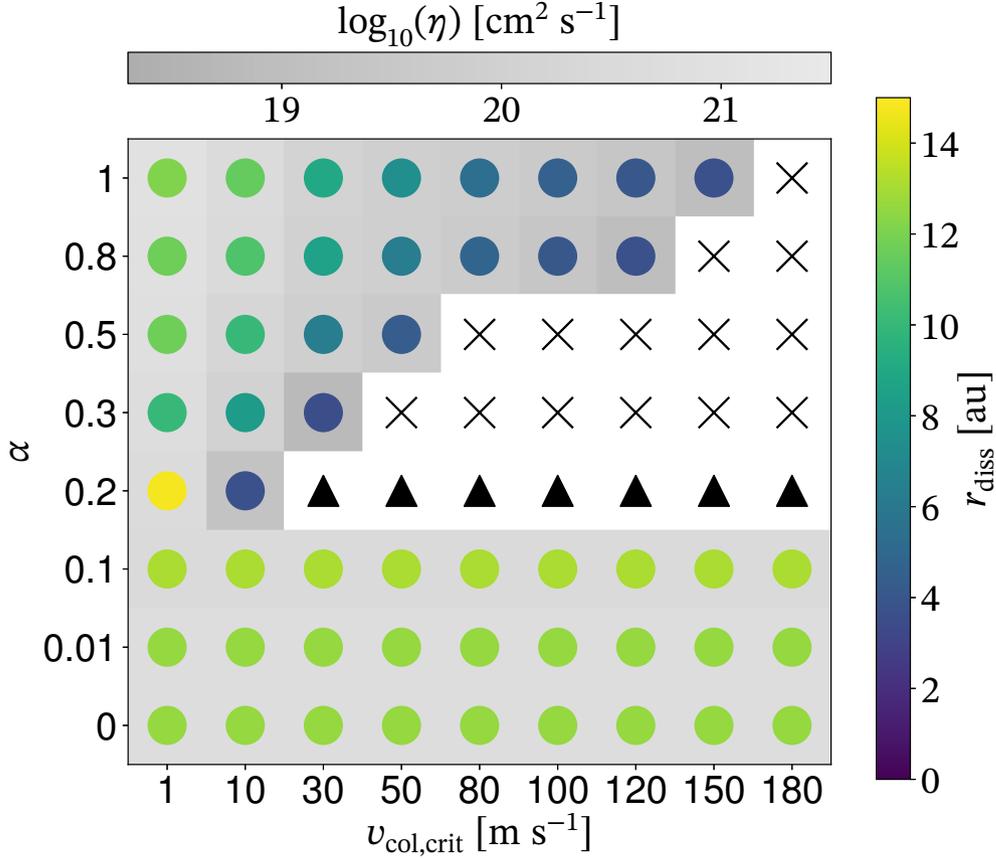}
  \caption{
Calculation results for $\alpha$--$v_{\mathrm{col, crit}}$ parameter plane, with monomer size fixed at  $a_{0}= 0.1\,  \mathrm{\mu m}$. The horizontal axis is $v_{\mathrm{crit, col}}$ and the vertical axis is $\alpha$. The circle symbols ($\bigcirc$) indicate models showing  $R_{\mathrm{m}} < 1$, while the cross symbols ($\times$) indicate models not showing  $R_{\mathrm{m}} < 1$. The triangle symbols ($\triangle$) indicate models transiently showing  $R_{\mathrm{m}} < 1$. 
The color of the circle symbol indicates the dissipation radius  corresponding to the Jeans length when the magnetic Reynolds number reaches $R_{\mathrm{m}} < 1$. 
Grey scale background for the models with circle symbol represents the magnitude of the maximum magnetic diffusion coefficient
 ( $\eta = {\rm max}(\eta_{\rm A}, \eta_{O}$)) when $R_{\mathrm{m}} < 1$. 
}
  \label{fig:a01}
\end{figure*}

We summarize the calculation results for the models with $a_{0}= 0.1\, \mathrm{\mu m}$ in Figure~\ref{fig:a01}. 
The figure represents whether non-ideal MHD effects (Ohmic dissipation and ambipolar diffusion) act in a high-density region  for 
each model with different $v_{\mathrm{crit, col}}$ and $\alpha$. 
In the figure, the circle symbols $(\circ )$ indicate cases where $R_{\mathrm{m}} < 1$ is realized for either Ohmic dissipation or ambipolar diffusion, or both,  the cross symbols $(\times )$ indicate cases where $R_{\mathrm{m}} < 1$ is not realized for both  Ohmic dissipation and ambipolar diffusion, and the triangle symbols ($\triangle$) indicate the cases where $R_{\mathrm{m}} < 1$ is transiently realized for either Ohmic dissipation or ambipolar diffusion, or both. 
The last case can be confirmed in the bottom panel in Figure~\ref{fig:a01_alpha02_v30_abund_eta} (for details, see \S\ref{sec:a01a02v30}).

The models with circle symbols have a gray scale background in the figure.
The background color is determined by the magnetic diffusion coefficient $\eta$ when the magnetic Reynolds number is lowest.
In addition,  for the models with the circle symbol, the dissipation length is represented by the color within the circle, in which the dissipation length corresponds to the Jeans length $L_{\mathrm{J}} = (1/2) \sqrt{\pi c_{s}^{2}/(G\rho_{g})}$ derived from the lowest density when $R_{\mathrm{m}} < 1$ is fulfilled.
We consider that the dissipation length shown in Figure~\ref{fig:a01} (and Figure~\ref{fig:a001}) is just an index for evaluating whether the disk can form in a high-density region. 
As shown in previous studies \citep[see review by][]{2022arXiv220913765T}, a rotationally supported disk tends to form in regions where magnetic dissipation effectively occurs. 
Figure~\ref{fig:a01} shows that the models with smaller dissipation lengths tend to have smaller magnetic diffusion coefficients.

For the models with relatively strong turbulence $\alpha \gtrsim 0.8$, $R_{\mathrm{m}} < 1$ is realized when the fragmentation velocity is smaller than  $v_{\mathrm{col, crit}} \lesssim 150\, \mathrm{m\,s^{-1}}$.
Thus, in a strongly turbulent environment,  dissipation of the magnetic field due to non-ideal MHD effects can act at a large fragmentation velocity ($v_{\mathrm{col, crit}} \simeq 150\, \mathrm{m\, s^{-1}}$).

When the turbulence intensity is moderate ($0.2 \lesssim \alpha \lesssim 0.5$), collisional fragmentation of dust particles is less likely to occur.
Thus, the dust particles coagulate and grow when $v_{\mathrm{col, crit}} \gtrsim  50$--$80\, \mathrm{m\,s^{-1}}$. 
In such cases, the magnetic diffusion coefficients are small.
The magnetic Reynolds number is $R_{\mathrm{m}}>1$ even in the high-density region because the decrease of the abundance (or total cross-section) of dust particles enhances the abundance of electrons and ions.
On the other hand, when the fragmentation velocity is smaller than  $v_{\mathrm{col, crit}} \lesssim 1$--$10\, \mathrm{m\, s^{-1}}$, $R_{\mathrm{m}}<1$ is realized due to the production of abundant small dust particles formed by collisional fragmentation.

In the range $0.2<\alpha<1$ (strong and moderate turbulence cases),  the turbulence intensity required for $R_{\mathrm{m}} < 1$ increases as the fragmentation velocity increases, as shown in  Figure~\ref{fig:a01}.
In addition, in such environments ($0.2<\alpha<1$),  a smaller fragmentation velocity can produce a larger disk (see the color within each circle)  because the production of small dust particles by collisional fragmentation becomes more efficient as the fragmentation velocity $v_{\mathrm{col, crit}}$ decreases. 
When the fragmentation velocity is as small as $v_{\mathrm{col, crit}} \lesssim 10 \ \mathrm{m\, s^{-1}}$, the dissipation length exceeds $r_{\mathrm{diss}} > 10 \ \mathrm{au}$. 
On the other hand,  when $v_{\mathrm{col, crit}}$ is larger than $120\, \mathrm{m\, s^{-1}}$, the dissipation length is smaller than $r_{\mathrm{diss}} \lesssim 4 \ \mathrm{au}$ in strongly turbulent environments ($\alpha> 0.8$)
\footnote{Again, we stress that the dissipation length here is just an index to discuss the ease of a disk formation. 
Three dimensional simulations are necessary to investigate the disk formation.}. 
In addition, for $0.2<\alpha<0.5$, it is expected that there is no magnetically inactive region when the fragmentation velocity is as large as $v_{\mathrm{col, crit}}>30$--$80\,\mathrm{m\, s^{-1}}$. 
In the magnetically active region, it is difficult to form the rotationally supported disk due to efficient angular momentum transport by magnetic braking.
Thus, a smaller fragmentation velocity is plausible for disk formation when the turbulence is moderate or strong.

Finally, we focus on cases with weak or zero turbulence ($\alpha \le 0.1$).
For these cases, the evolution of the dust particle size distribution is almost the same as described in 
\S\ref{sec:a01a0v1} because dust growth due to turbulence is not promoted. 
The magnetic diffusion coefficients follow almost the same path, and the magnetic Reynolds number falls below unity in high-density regions. 
In addition, the density for which $R<1$ is realized is almost the same among the models with $\alpha \le 0.1$ (see, Fig.~\ref{fig:a01_alpha0_v1_abund_eta}).
Thus, the dissipation length is also the same and $r_{\mathrm{diss}} \simeq 13$\,au for these models (Fig.~\ref{fig:a01}).

\subsection{Model of monomer size of $a_{0} = 0.01\, \mathrm{\mu m}$}
\label{sec:a001}

%%%%%%
% Fig. 12
%%%%%%
\begin{figure}
  \centering
  \includegraphics[width=80mm]{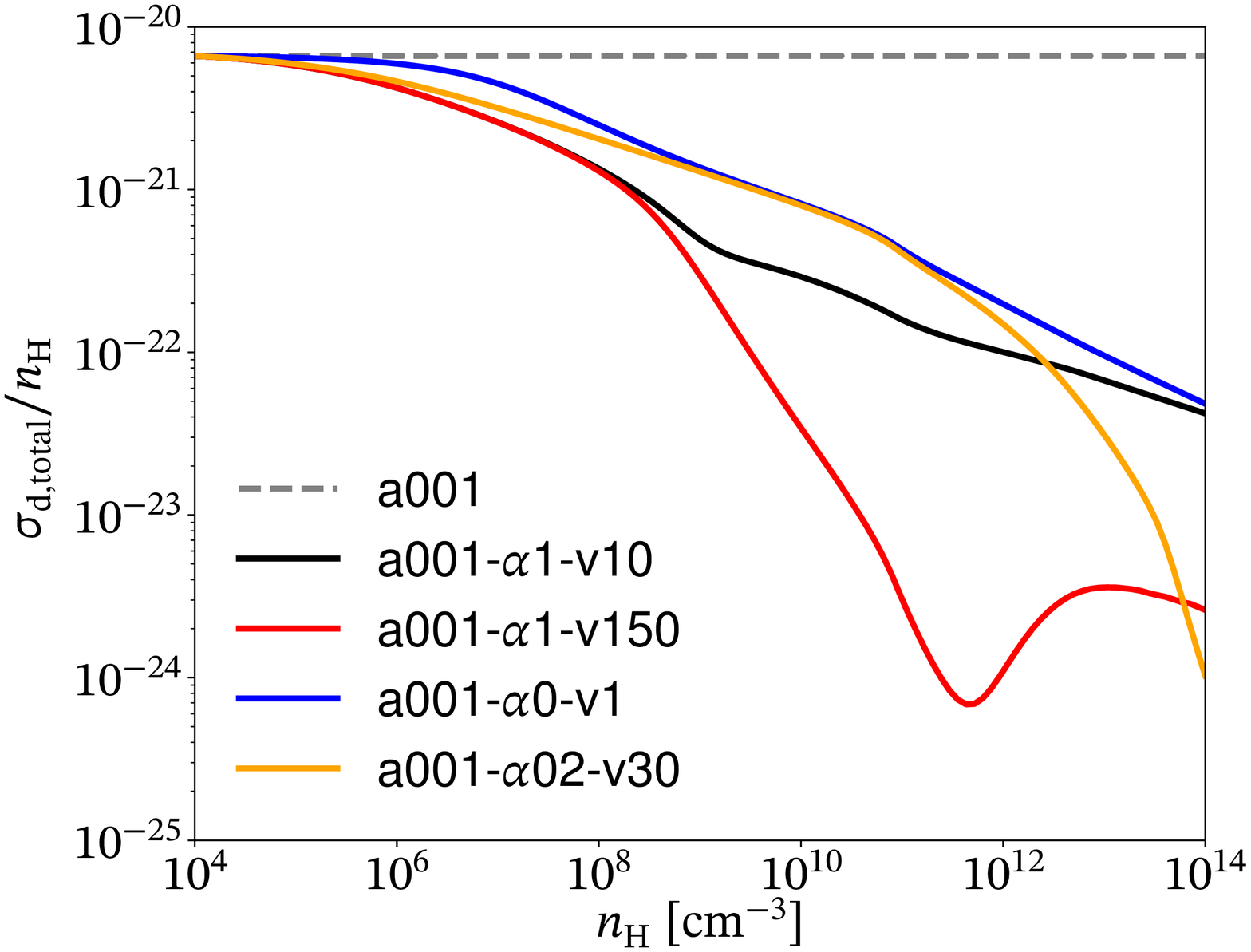}
  \caption{Same as Figure~\ref{fig:a01_dust_cross_section} but for the case of monomer size $a_{0} = 0.01\, \mathrm{\mu m}$.}
  \label{fig:a001_dust_cross_section}
\end{figure}

%%%%%%
% Fig. 13
%%%%%%
\begin{figure*}
  \centering
  \includegraphics[width=160mm]{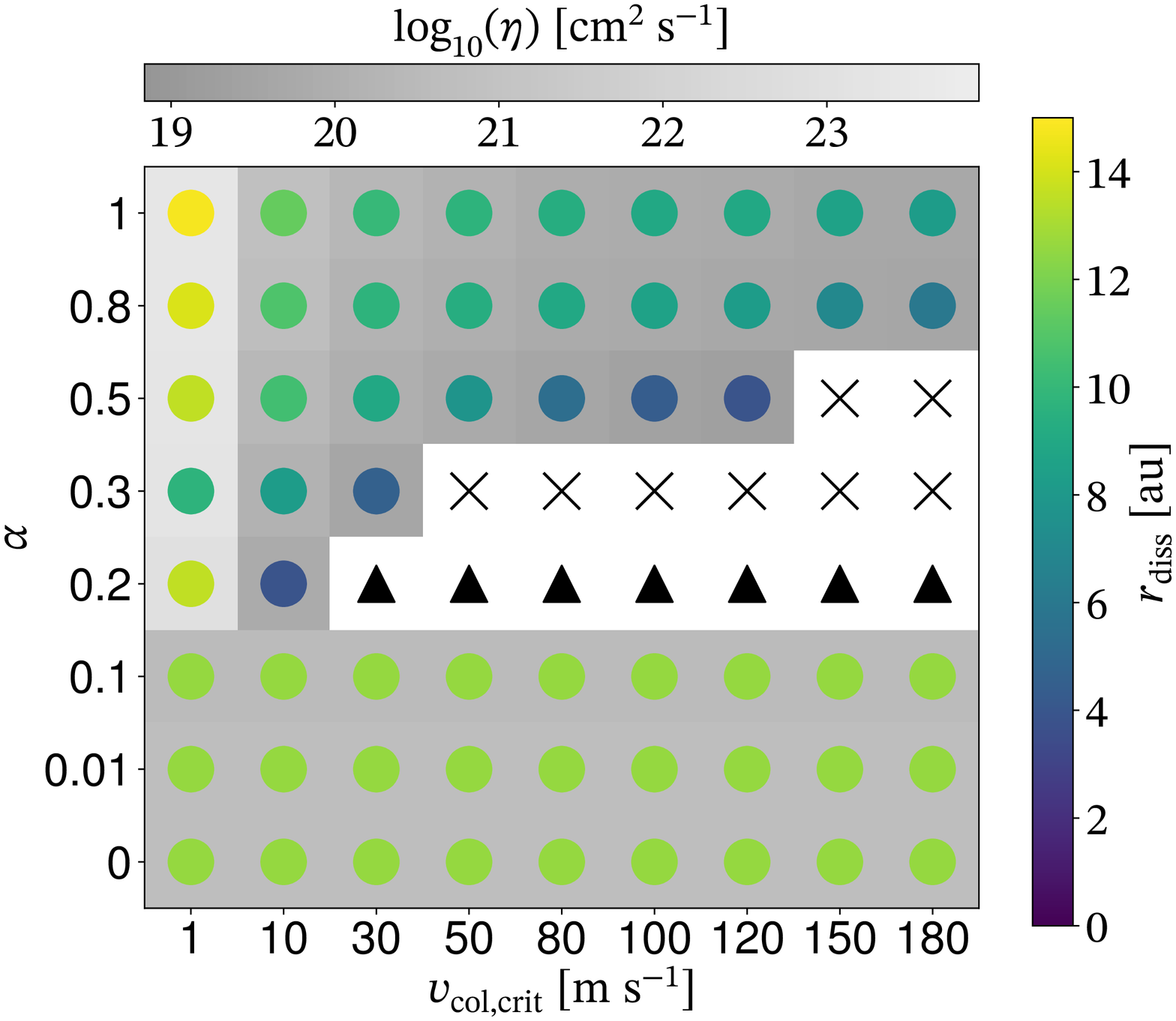}
  \caption{Same as Figure~\ref{fig:a01} but for the case of monomer size $a_{0} = 0.01\, \mathrm{\mu m}$.}
  \label{fig:a001}
\end{figure*}

We describe the results for models with a monomer size of $a_{0}=0.01\, \mathrm{\mu m}$. 
Figure~\ref{fig:a001_dust_cross_section} shows the evolution of the total dust cross section for the case of $a_{0} = 0.01 \ \rm{\mu m}$.
The turbulence intensity and the fragmentation velocity for each model in Figure~\ref{fig:a001_dust_cross_section} are the same as for the model with $a_{0} = 0.1 \ \rm{\mu m}$. 

Comparing Figure~\ref{fig:a001_dust_cross_section} with Figure~\ref{fig:a01_dust_cross_section}, it can be seen that the evolution of the total dust cross section for models with $a_{0} = 0.01 \ \rm{\mu m}$ is roughly the same as for models with $a_{0} = 0.1 \ \rm{\mu m}$.
However, there are subtle differences. 
Brownian motion dominates turbulence for dust collisions when the dust particles are smaller than  $a \simeq 0.1 \ \rm{\mu m}$. 
The relative velocity between dust particles due to Brownian motion increases as the dust particle size (or dust mass) decreases.
Thus, coagulation growth by Brownian motion is promoted for smaller dust particles.  
Therefore, the density at which the total dust cross section begins to decrease due to coagulation growth is lower for the model with $a_{0}=0.01\, \mathrm{\mu m}$ than for the model with $a = 0.1 \ \rm{\mu m}$.
However, the total dust cross section for the model with $a_{0}=0.01\, \mathrm{\mu m}$ is larger than for the model with $a_{0}=0.1\, \mathrm{\mu m}$.
This is because dust particles are more abundant for a smaller dust particle size when the total dust mass is the same.
Therefore, the amount of adsorption of charged particles onto the dust increases with decreasing size of the smallest dust particles; as a result, the degree of ionization also decreases. 

Figure~\ref{fig:a001} is the same as Figure~\ref{fig:a01} but for the cases with a monomer size of $a_{0}=0.01\, \mathrm{\mu m}$. 
For strong turbulence cases with $\alpha \gtrsim 0.8$, $R_{\mathrm{m}} < 1$ is realized for all ranges of $v_{\mathrm{col, crit}}$. 
Thus, for these models, disk formation is expected to be independent of the fragmentation velocity  adopted in this study. 
However, the magnetic diffusion coefficient (gray scale background) increases as the fragmentation velocity decreases. 
Thus, the disk formation is more plausible for the models with a smaller fragmentation velocity.
In addition, the dissipation length increases as the fragmentation velocity decreases, as also seen in Figure~\ref{fig:a01}.

The dissipation length is larger for models with $a_{0} = 0.01\, \mathrm{\mu m}$ (Fig.~\ref{fig:a001}) than for models with $a_{0} = 0.1\, \mathrm{\mu m}$ (Fig.~\ref{fig:a01}).
This is because the dust total cross section increases with decreasing smallest dust particle size, as explained above.
Thus, for models with a small monomer size $a_{0} = 0.01\, \mathrm{\mu m}$, the magnetic diffusion coefficients increase and strong non-ideal MHD effects occur.
Therefore, the condition  $R_{\mathrm{m}} < 1$ tends to be realized in high-density regions.
Unlike for models with $a_{0} = 0.1\, \mathrm{\mu m}$,  disk formation in the magnetically inactive region is expected as long as the turbulence intensity is high ($\alpha>0.5$) or low ($\alpha<0.2$) even when the fragmentation velocity is as large as $v_{\mathrm{col, crit}} = 180 \ \mathrm{m\, s^{-1}}$ for models with $a_{0} = 0.01\, \mathrm{\mu m}$.

In the case of moderate turbulence  $\alpha = 0.2$--$0.5$, the tendency of the non-ideal MHD effects and the dissipation length and magnetic diffusion coefficient seen in Figure~\ref{fig:a001} is almost the same as in Figure~\ref{fig:a01}, 
while disk formation seems to be more plausible for models with $a_{0} = 0.01 \ \mathrm{\mu m}$ than for models with $a_{0} = 0.1 \ \mathrm{\mu m}$. 
When the parameters $\alpha$ and $v_{\mathrm{col, crit}}$ are the same, the dissipation length and magnetic diffusion coefficient tend to be larger  in models with $a_{0} = 0.01\, \mathrm{\mu m}$ than in models with $a_{0} = 0.1\, \mathrm{\mu m}$. 

In the weak turbulence case $\alpha\lesssim 0.1$, the results are almost the same for both models.  
Thus, it is expected that disks form independent of $v_{\mathrm{col, crit}}$ and $a_{0}$ in weak turbulence environments. 

\section{Discussion}
\label{se:discussion}
\subsection{Magnetic braking catastrophe}
Non-ideal MHD effects are considered to be key for resolving the magnetic braking catastrophe that prevents the formation of circumstellar disks  due to excessive angular momentum transport by magnetic braking  \citep{2003ApJ...599..363A,2008ApJ...681.1356M}.
Non-ideal MHD effects, Ohmic diffusion and ambipolar diffusion, can remove the magnetic field from the central region and weaken the efficiency of magnetic braking. 
Thus, they promote the formation of a circumstellar disk \citep{2010A&A...521L..56D, 2010ApJ...724.1006M, 2012A&A...541A..35D, 2013ApJ...763....6T,2015ApJ...801..117T, 2015MNRAS.452..278T,2021MNRAS.502.4911X,2021MNRAS.508.2142X}.

This study showed that dissipation of the magnetic field due to non-ideal MHD effects does not work efficiently in moderately turbulent environments ($\alpha \simeq 0.2$--$0.5$) as long as the fragmentation velocity is larger than  $v_{\mathrm{col, crit}} \gtrsim 50$--$80 \ \mathrm{m\,s^{-1}}$. 
Even in the case of strong turbulence ($\alpha \gtrsim 0.8$), non-ideal MHD effects are not activated when the fragmentation velocity of the dust particles is larger than $v_{\mathrm{col, crit}} \gtrsim 150\, \mathrm{m\,s^{-1}}$ (e.g. a01-$\alpha$1-v180).
Thus, in terms of disk formation, the dust fragmentation velocity can be severely limited, which is also important for dust growth and further planet formation. 

In the case of weak turbulence ($\alpha < 0.2$), non-ideal MHD effects can remove the magnetic field from a high-density region regardless of the fragmentation velocity. 
Thus, non-ideal MHD effects should reliably contribute to disk formation.

The evolution of the dust particle size distribution significantly depends on the turbulence velocity and the fragmentation velocity of the dust particles.
The dissipation efficiency of the magnetic field due to non-ideal MHD effects strongly depends on the dust particle size distribution, which should influence the disk formation process.
However, our calculations are based on a one-zone model.
Thus, three dimensional non-ideal MHD simulations with a multi-fluid of gas and dust \citep{2017MNRAS.465.1089B, 2019A&A...626A..96L,2020A&A...641A.112L,2021ApJ...913..148T,2021ApJ...920L..35T, 2022MNRAS.515.6073K, 2022MNRAS.tmp.3298K} are necessary to clarify the star and disk formation processes in more detail. 
Recently, \citet{2023A&A...670A..61M} performed three-dimensional non-ideal MHD simulations considering dust growth. 
They showed that dust growth could weaken  the non-ideal MHD effects in high-density regions (or disk-forming regions), 
affecting the formation and evolution of the circumstellar disk.
Their results indicate that dust growth is closely related to disk growth.

\subsection{Fragmentation velocity of dust particles}
As described in \S\ref{sec:fragmentation_model}, the dust fragmentation model adopted in this study is based on a numerical calculations of dust collisions.
Some numerical studies \citep[e.g.][]{2013A&A...559A..62W} have suggested that the fragmentation velocity of dust particles composed of  silicate and $\mathrm{H_{2}O}$  can be represented  by, respectively,
\begin{equation}
  v_{\mathrm{col, crit}} = 1 \textrm{--} 10 \left(\frac{a_{0}}{0.1\,\mu {\rm m}}\right)^{-5/6}\,  \mathrm{m\,s^{-1}} \ \text{for silicate},
  \label{eq:frag_vel_silicate}
\end{equation}
and
\begin{equation}
  v_{\mathrm{col, crit}} = 50 \textrm{--} 80 \left(\frac{a_{0}}{0.1\, \mu {\rm m}}\right)^{-5/6}\, \mathrm{m\,s^{-1}} \ \text{for $\mathrm{H_{2}O}$ ice}, 
  \label{eq:frag_vel_H2Oice}
\end{equation}
where $a_{0}$ is the monomer size adopted in the dust fragmentation simulations. 
In the following, we assume that $a_{0}$ can be related to $v_{\mathrm{col, crit}}$ through equations~(\ref{eq:frag_vel_silicate}) and (\ref{eq:frag_vel_H2Oice}).  

First, we discuss models having a monomer size of $a_{0} = 0.1\, \mathrm{\mu m}$. 
With $a_{0} = 0.1\, \mathrm{\mu m}$, equations (\ref{eq:frag_vel_silicate}) and (\ref{eq:frag_vel_H2Oice}) give $v_{\mathrm{col, crit}} = 1$--$10\, \mathrm{m\,s^{-1}}$ for silicate dust and $v_{\mathrm{col, crit}} = 50$--$80\, \mathrm{m\,s^{-1}}$ for $\mathrm{H_{2}O}$ ice.
When the dust is composed of $\mathrm{H_{2}O}$ ice particles, non-ideal MHD effects are not strong for a moderate turbulence intensity of $0.2$--$0.5$ times the speed of sound ($\alpha=0.2$--$0.5$, Fig.~\ref{fig:a01}) because collisional fragmentation does not occur frequently.
On the other hand, when the turbulence is weak, $\alpha \lesssim 0.1$, or strong, $\alpha \gtrsim 0.8$, non-ideal MHD effects are effective in high-density regions
in the range $v_{\mathrm{col, crit}} = 50$--$80 \ \rm{m\,s^{-1}}$ (Fig.~\ref{fig:a01}).
When the dust is composed of silicate particles, collisional fragmentation can occur even with a moderate turbulence intensity,  $\alpha=0.2$--$0.5$.
As a result, with silicate dust, non-ideal MHD effects become effective in high-density regions regardless of the turbulence intensity (Fig.~\ref{fig:a01}).

Next, we consider a monomer size of $a_{0} = 0.01\, \mathrm{\mu m}$.
With $a_{0} = 0.01\, \mathrm{\mu m}$,  equations (\ref{eq:frag_vel_silicate}) and (\ref{eq:frag_vel_H2Oice}) give $v_{\mathrm{col, crit}} = 6.8$--$68\, \mathrm{m\,s^{-1}}$ for silicate  and $v_{\mathrm{col, crit}} = 340$--$545\, \mathrm{m\,s^{-1}}$ for $\mathrm{H_{2}O}$ ice.
Thus, the fragmentation velocity  is higher for $a_{0} = 0.01\, \mathrm{\mu m}$  than for $a_{0} = 0.1\, \mathrm{\mu m}$ 
because the binding energy between dust particles increase as the monomer size of the dust particles decreases.
In the case of silicate dust, non-ideal MHD effects occur regardless of the turbulence intensity, as in the case of $a_{0} = 0.1\, \mathrm{\mu m}$. 
In the case of $\mathrm{H_{2}O}$ ice dust,  the fragmentation velocity derived from equation (\ref{eq:frag_vel_H2Oice}) 
is out of the range set in this study.
However, it is expected that the density range where non-ideal MHD effects occur is narrow and almost the same as in the case of $v_{\mathrm{col, crit}} = 180\, \mathrm{m\,s^{-1}}$ because fragmentation is less likely to occur. 
In this case, the mass of small dust particles produced due to fragmentation decreases as the fragmentation velocity increases. 

We separately considered silicate and $\mathrm{H_{2}O}$ ice as the dust composition in the above. 
In other words, we assumed that the dust is composed of either silicate or $\mathrm{H_{2}O}$ ice particles. 
However, in reality, dust particles  are considered to have an internal structure of a central core and a mantle. The central silicate core is enclosed by a mantle  composed of $\mathrm{H_{2}O}$ \citep{2015ARA&A..53..541B}.

When the gas temperature  reaches $T \simeq 150 \mathrm{K}$, the $\mathrm{H_{2}O}$ ice mantle on the dust particle surface sublimates. 
The silicate core does not evaporate at that temperature.
Thus, for disk formation in a collapsing cloud core, the fragmentation velocity should be large at relatively low gas densities because the $\mathrm{H_{2}O}$ ice mantle contributes to dust particle growth. 
Thus, coagulation growth of dust particles proceeds without fragmentation unless the turbulence is very weak ($\alpha<0.2$) or strong ($\alpha>0.8$). 

As the gas density increases in the collapsing cloud, the fragmentation velocity should be small because the $\mathrm{H_{2}O}$ ice mantle evaporates and the silicate core is exposed. 
Then,  dust particles easily fragment, as in the case of silicate dust.
As a result, non-ideal MHD effects can contribute to dissipation of the magnetic field because the fragmentation velocity for the silicate core is low and small dust particles are produced. 
Thus, we need to understand how the composition of the dust particle surface changes in order to investigate the evolution of the dust particle size distribution more realistically.

\subsection{Turbulence within a collapsing cloud core} 
We showed that the evolution of the dust particle size distribution strongly depends on the turbulence strength in the collapsing cloud core.
When the relative velocity between dust particles due to turbulence is large,
the dust particles frequently collide and are likely to fragment.
Since the turbulence velocity in molecular cloud cores is not well understood, we treated it as a parameter such that the ratio of the turbulence velocity to the speed of sound is constant. 

Recently, \citet{2021ApJ...915..107H} have shown that subsonic turbulence can be amplified to be comparable to the speed of sound in a collapsing cloud core. 
They performed calculations with a single polytopic index in the equation of state because they assumed a primordial environment.
In present-day star formation, the polytopic index changes with increasing central density \citep{2012PTEP.2012aA307I}. 
Therefore, we cannot easily apply their calculation results.
However, to investigate the dust particle size distribution in collapsing cores more precisely, we may need to determine both the evolution of the dust particle size distribution and the evolution of turbulence simultaneously. 

\subsection{Comparison with other studies}

In this section, we compare our results with previous studies on the evolution of the dust particle size distribution during the core-collapse phase.

\citet{2009MNRAS.399.1795H} calculated the evolution of the dust particle size distribution by considering only Brownian motion.
Their results are consistent with those for our model without turbulence (only Brownian motion) as described in \S\ref{sec:a01a02v30}.

\citet{2023MNRAS.518.3326L} calculated the dust particle size distribution assuming that the injection velocity for turbulence was equal to the speed of sound, as in Paper I and the model with $\alpha=1$ in the present study.
They used a model in which the fragmentation velocity depends on the mass ratio of the colliding dust particles.
When particles with the same mass collide, the fragmentation velocity for their model is about $15 \ \mathrm{m\, s^{-1}}$ for silicate dust and about $300 \ \mathrm{m\, s^{-1}}$ for $\mathrm{H_{2}O}$ ice dust.
On the other hand, the present study and that reported in Paper I used a constant fragmentation velocity for all mass ratios of the colliding dust particles.
The fragmentation velocity can depend on the mass ratio of the colliding dust particles \citep[e.g.][]{2021ApJ...915...22H}.
However, the detailed dependence of the fragmentation velocity on the dust mass ratio is not well understood.
Thus, many dust growth study calculations have used a constant fragmentation velocity that is independent of the mass ratio of the colliding dust particles 
\citep[e.g.][]{2010A&A...513A..79B}.
Although the model for fragmentation velocity is different, dust particle size evolution in the central region of the collapsing core is similar to the results for our models with
$v_{\mathrm{col, crit}} \simeq 10$--$30 \ \mathrm{m\, s^{-1}}$ and $v_{\mathrm{col, crit}} \simeq 180 \ \mathrm{m\, s^{-1}}$.
This means that the magnetic diffusion coefficients in the central region have a similar evolution in our study and \citet{2023MNRAS.518.3326L}.

Although we calculated the charge state of the dust particles, we did not consider the effect on dust particle motion, for example ambipolar diffusion drift.
\citet{2020A&A...643A..17G} and \citet{2023MNRAS.518.3326L} showed that ambipolar diffusion can remove small dust particles $a \lesssim 0.1 \ \mathrm{\mu m}$ in a low-density region.
If we calculate the dust particle size evolution including ambipolar diffusion drift of dust particles for a model with a monomer size of $a = 0.01 \ \mathrm{\mu m}$,
the total dust cross section may decrease compared to the results described in \S\ref{sec:a001} (Fig.~\ref{fig:a001_dust_cross_section}).
We will investigate the evolution of the dust particle size distribution, taking into account the dust charging effect, in a future study.

\section{summary}
\label{sec:summary}
We  investigated how the dust particle size evolution can affect non-ideal MHD effects in a collapsing core, especially in terms of the turbulence intensity
and the fragmentation velocity for dust particles. 
Considering two different dust monomer sizes, we calculated the dust particle size evolution  using the three parameters of turbulence intensity, 
fragmentation velocity, and monomer size. 
The calculation procedure is almost the same as in Paper I. 
We also evaluated the magnetic diffusion coefficients and magnetic Reynolds number for each model with different parameters 
to constrain the parameters of the turbulence intensity and fragmentation velocity in terms of circumstellar disk formation. 

When the turbulence intensity is smaller than $0.1$ times the speed of sound $(\alpha \lesssim 0.1)$, the turbulence has little effect on the dust particle size evolution.
In this case, the dust particles do not grow significantly regardless of the fragmentation velocity, and non-ideal MHD effects can contribute to the dissipation of the magnetic field, as in the case where the dust particle size distribution is not considered.

When the turbulence intensity is stronger than $0.8$ times the speed of sound $(\alpha \gtrsim 0.8)$, small dust particles can be produced by fragmentation even for relatively
high fragmentation velocities ($v_{\mathrm{col, crit}} \simeq 150$--$180 \ \mathrm{m\, s^{-1}}$).
The abundant small dust particles allow for efficient adsorption of charged particles (especially electrons) on the dust particle surfaces, 
resulting in a decrease in the abundance of charged particles and an enhancement of the magnetic diffusion coefficients at high density.
In this case, the magnetic Reynolds number drops below unity in high-density regions and non-ideal MHD effects are significant.
The density at which the magnetic Reynolds number is below unity increases as the fragmentation velocity increases.
Thus, a large fragmentation velocity for dust particles can result in the formation of a relatively small disk.

When the turbulence intensity is moderate, $ \ 0.2 \lesssim \alpha \lesssim 0.5$, a fragmentation velocity less than $30$--$50\, \rm{m\, s^{-1}}$ 
is necessary to activate non-ideal MHD effects in high-density regions. 
On the other hand, when the fragmentation velocity is $\gtrsim 50 \ \rm{m\, s^{-1}}$, dust particles coagulate and grow without fragmentation, 
reducing the abundance  or total cross section of dust particles. 
Thus, non-ideal MHD effects cannot contribute to the dissipation of the magnetic field.

We discuss our results for the fragmentation velocity obtained from numerical dust collision calculations in the cases of dust composed of silicate or $\mathrm{H_{2}O}$ ice.
The fragmentation velocity varies with the dust particle surface composition 
and the difference affects the dust particle size evolution and the magnetic diffusion coefficients.
Therefore, it is important to understand how the composition of the dust (especially its surface) changes during the disk formation stage in star-forming cores.

\section*{Acknowledgements}
This work was supported by the Japan Society for the Promotion of Science KAKENHI (JP17H06360, JP17K05387, JP17KK0096, JP21H00046, JP21K03617: MNM), 
JSPS KAKENHI grants (JP22J11129: YK), and NAOJ ALMA Scientific Research Grant Code 2022-22B.

% The Acknowledgements section is not numbered. Here you can thank helpful
% colleagues, acknowledge funding agencies, telescopes and facilities used etc.

%%%%%%%%%%%%%%%%%%%%%%%%%%%%%%%%%%%%%%%%%%%%%%%%%%
\section*{Data Availability}
The data underlying this article are available on request.
 
% The inclusion of a Data Availability Statement is a requirement for articles published in MNRAS. Data Availability Statements provide a standardised format for readers to understand the availability of data underlying the research results described in the article. The statement may refer to original data generated in the course of the study or to third-party data analysed in the article. The statement should describe and provide means of access, where possible, by linking to the data or providing the required accession numbers for the relevant databases or DOIs.



\bibliographystyle{mnras}
\bibliography{example} % if your bibtex file is called example.bib

%\begin{thebibliography}{99}
%\end{thebibliography}

%%%%%%%%%%%%%%%%%%%%%%%%%%%%%%%%%%%%%%%%%%%%%%%%%%

%%%%%%%%%%%%%%%%% APPENDICES %%%%%%%%%%%%%%%%%%%%%

% \appendix

% \section{Some extra material}

% If you want to present additional material which would interrupt the flow of the main paper,
% it can be placed in an Appendix which appears after the list of references.

%%%%%%%%%%%%%%%%%%%%%%%%%%%%%%%%%%%%%%%%%%%%%%%%%%

% Don't change these lines
\bsp	% typesetting comment
\label{lastpage}
\end{document}